\input harvmac
\noblackbox
\newcount\figno
\figno=0
\def\fig#1#2#3{
\par\begingroup\parindent=0pt\leftskip=1cm\rightskip=1cm\parindent=0pt
\baselineskip=11pt
\global\advance\figno by 1
\midinsert
\epsfxsize=#3
\centerline{\epsfbox{#2}}
\vskip 12pt
\centerline{{\bf Figure \the\figno:} #1}\par
\endinsert\endgroup\par}
\def\figlabel#1{\xdef#1{\the\figno}}

\def\np#1#2#3{Nucl. Phys. {\bf B#1} (#2) #3}
\def\pl#1#2#3{Phys. Lett. {\bf B#1} (#2) #3}
\def\prl#1#2#3{Phys. Rev. Lett. {\bf #1} (#2) #3}
\def\prd#1#2#3{Phys. Rev. {\bf D#1} (#2) #3}

\def\cmp#1#2#3{Comm. Math. Phys. {\bf #1} (#2) #3}


\font\cmss=cmss10
\font\cmsss=cmss10 at 7pt
\def\rlx{\relax\leavevmode}
\def\inbar{\vrule height1.5ex width.4pt depth0pt}
\def\IC{\relax\,\hbox{$\inbar\kern-.3em{\rm C}$}}
\def\IN{\relax{\rm I\kern-.18em N}}
\def\IP{\relax{\rm I\kern-.18em P}}
\def\ZZ{\rlx\leavevmode\ifmmode\mathchoice{\hbox{\cmss Z\kern-.4em Z}}
 {\hbox{\cmss Z\kern-.4em Z}}{\lower.9pt\hbox{\cmsss Z\kern-.36em Z}}
 {\lower1.2pt\hbox{\cmsss Z\kern-.36em Z}}\else{\cmss Z\kern-.4em
 Z}\fi}
\def\IZ{\relax\ifmmode\mathchoice
{\hbox{\cmss Z\kern-.4em Z}}{\hbox{\cmss Z\kern-.4em Z}}
{\lower.9pt\hbox{\cmsss Z\kern-.4em Z}}
{\lower1.2pt\hbox{\cmsss Z\kern-.4em Z}}\else{\cmss Z\kern-.4em
Z}\fi}

\def\narrowplus{\kern -.04truein + \kern -.03truein}
\def\narrowminus{- \kern -.04truein}
\def\narrowminussub{\kern -.02truein - \kern -.01truein}

\def\b{{\beta}}
\def\a{{\alpha}}
\def\g{{\gamma}}
\def\gg{{\tilde \gamma}}

\def\l{{\lambda}}

\def\r{{\rightarrow}}

\def\frac#1#2{{#1\over #2}}

\def\com#1#2{{ \left[ #1, #2 \right] }}

\def\acom#1#2{{ \left\{ #1, #2 \right\} }}

\def\IZ{\relax\ifmmode\mathchoice
{\hbox{\cmss Z\kern-.4em Z}}{\hbox{\cmss Z\kern-.4em Z}}
{\lower.9pt\hbox{\cmsss Z\kern-.4em Z}}
{\lower1.2pt\hbox{\cmsss Z\kern-.4em Z}}\else{\cmss Z\kern-.4em
Z}\fi}
\def\IB{\relax{\rm I\kern-.18em B}}
\def\IC{{\relax\hbox{$\inbar\kern-.3em{\rm C}$}}}
\def\ID{\relax{\rm I\kern-.18em D}}
\def\IE{\relax{\rm I\kern-.18em E}}
\def\IF{\relax{\rm I\kern-.18em F}}
\def\IG{\relax\hbox{$\inbar\kern-.3em{\rm G}$}}
\def\IGa{\relax\hbox{${\rm I}\kern-.18em\Gamma$}}
\def\IH{\relax{\rm I\kern-.18em H}}
\def\II{\relax{\rm I\kern-.18em I}}
\def\IK{\relax{\rm I\kern-.18em K}}
\def\IP{\relax{\rm I\kern-.18em P}}

\font\cmss=cmss10 \font\cmsss=cmss10 at 7pt
\def\IR{\relax{\rm I\kern-.18em R}}

\def\f{\psi}
\def\l{\lambda}

\def\s{{\sigma}}
\def\ss{{\tilde s}}
\def\1{{\bf 1}}
\def\3{{\bf 3}}
\def\4{{\bf 4}}
\def\5{{\bf 5}}
\def\7{{\bf 7}}
\def\2{{\bf 2}}
\def\8{{\bf 8}}

%

%
%
\def\eqnn#1{\xdef #1{(\secsym\the\meqno)}\writedef{#1\leftbracket#1}%
\global\advance\meqno by1\wrlabeL#1}
\def\eqna#1{\xdef #1##1{\hbox{$(\secsym\the\meqno##1)$}}
\writedef{#1\numbersign1\leftbracket#1{\numbersign1}}%
\global\advance\meqno by1\wrlabeL{#1$\{\}$}}
\def\eqn#1#2{\xdef #1{(\secsym\the\meqno)}\writedef{#1\leftbracket#1}%
\global\advance\meqno by1$$#2\eqno#1\eqlabeL#1$$}



\lref\rsegal{G. Segal and A. Selby, \cmp{177}{1996}{775}. }
\lref\rdorold{N. Dorey, V. Khoze, M. Mattis, D. Tong and S. Vandoren, 
hep-th/9703228, \np{502}{1997}{59}.}
\lref\rsen{A. Sen, hep-th/9402002, Int. J. Mod. Phys. {\bf A9} (1994) 3707;
hep-th/9402032, \pl{329}{1994}{217}. }
\lref\rcallias{C. Callias, \cmp{62}{1978}{213}\semi E. Weinberg, 
\prd{20}{1979}{936}.}
\lref\rblum{J. Blum, hep-th/9401133, \pl{333}{1994}{92}.}
\lref\rsgreen{M. B. Green and S. Sethi, hep-th/9808061.}
\lref\rbateman{H. Bateman, ed. A. Erdelyi, 
{\it Higher Transcendental Functions, vol. 2}, McGraw-Hill Book Company, 1953. }
\lref\ryi{P. Yi, hep-th/9704098, \np{505}{1997}{307}.}
\lref\rsavmark{S. Sethi and M. Stern, hep-th/9705046, \cmp{194}{1998}{675}.}
\lref\rwitthmon{E. Witten, hep-th/9511030, \np{460}{1996}{541}.}
\lref\rgg{M. B. Green and M. Gutperle, hep-th/9711107, JHEP 01 (1998) 05.}
\lref\rgreg{G. Moore, N. Nekrasov and S. Shatashvili, hep-th/9803265.}
\lref\rSS{S. Sethi and L. Susskind, hep-th/9702101,
    \pl{400}{1997}{265}.}
\lref\rBS{T. Banks and N. Seiberg,
    hep-th/9702187, \np{497}{1997}{41}.}
\lref\rreview{N. Seiberg, hep-th/9705117.}
\lref\rpp{J. Polchinski and P. Pouliot, hep-th/9704029, \prd{56}{1997}{6601}.}
\lref\rds{M. Dine and N. Seiberg, hep-th/9705057, \pl{409}{1997}{239}.}
\lref\rdorey{N. Dorey, V. Khoze and M. Mattis, hep-th/9704197, 
\np{502}{1997}{94}.}
\lref\rbf{T. Banks, W. Fischler, N. Seiberg and L. Susskind, hep-th/9705190, 
\pl{408}{1997}{111}.}
\lref\rlowe{D. Lowe, hep-th/9810075.}
\lref\rhoppe{G. M. Graf and J. Hoppe, hep-th/9805080.}

\lref\rK{N. Ishibashi, H. Kawai, Y. Kitazawa and A. Tsuchiya, hep-th/9612115.}
\lref\rCallias{C. Callias, Commun. Math. Phys. {\bf 62} (1978), 213.}
\lref\rPD{J. Polchinski, hep-th/9510017, \prl{\bf 75}{1995}{47}.}
\lref\rWDB{E. Witten,  hep-th/9510135, Nucl. Phys. {\bf B460} (1996) 335.}
\lref\rSSZ{S. Sethi, M. Stern, and E. Zaslow, Nucl. Phys. {\bf B457} (1995)
484.}
\lref\rGH{J. Gauntlett and J. Harvey, Nucl. Phys. {\bf B463} 287. }
\lref\rAS{A. Sen, Phys. Rev. {\bf D53} (1996) 2874; Phys. Rev. {\bf D54} (1996)
2964.}
\lref\rWI{E. Witten, Nucl. Phys. {\bf B202} (1982) 253.}
\lref\rPKT{P. K. Townsend, Phys. Lett. {\bf B350} (1995) 184.}
\lref\rWSD{E. Witten, Nucl. Phys. {\bf B443} (1995) 85.}
\lref\rASS{A. Strominger, Nucl. Phys. {\bf B451} (1995) 96.}
\lref\rBSV{M. Bershadsky, V. Sadov, and C. Vafa, Nucl. Phys. {\bf B463}
(1996) 420.}
\lref\rBSS{L. Brink, J. H. Schwarz and J. Scherk, Nucl. Phys. {\bf B121}
(1977) 77.}
\lref\rCH{M. Claudson and M. Halpern, Nucl. Phys. {\bf B250} (1985) 689.}
\lref\rSM{B. Simon, Ann. Phys. {\bf 146} (1983), 209.}
\lref\rGJ{J. Glimm and A. Jaffe, {\sl Quantum Physics, A Functional Integral
Point of View},
Springer-Verlag (New York), 1981.}
\lref\rADD{ U. H. Danielsson, G. Ferretti, B. Sundborg, Int. J. Mod. Phys. {\bf
A11} (1996) 5463\semi   D. Kabat and P. Pouliot, Phys. Rev. Lett. {\bf 77}
(1996), 1004.}
\lref\rDKPS{ M. R. Douglas, D. Kabat, P. Pouliot and S. Shenker,
hep-th/9608024,
Nucl. Phys. {\bf B485} (1997), 85.}
\lref\rhmon{S. Sethi and M. Stern, hep-th/9607145, Phys. Lett. {\bf B398} 
(1997), 47.}
\lref\rBFSS{T. Banks, W. Fischler, S. H. Shenker, and L. Susskind, 
hep-th/9610043,
Phys. Rev. {\bf D55} (1997) 5112.}
\lref\rBHN{ B. de Wit, J. Hoppe and H. Nicolai, Nucl. Phys. {\bf B305}
(1988), 545\semi
B. de Wit, M. M. Luscher, and H. Nicolai, Nucl. Phys. {\bf B320} (1989),
135\semi
B. de Wit, V. Marquard, and H. Nicolai, Comm. Math. Phys. {\bf 128} (1990),
39.}
\lref\rT{ P. Townsend, Phys. Lett. {\bf B373} (1996) 68.}
\lref\rLS{L. Susskind, hep-th/9704080.}
\lref\rFH{J. Frohlich and J. Hoppe, hep-th/9701119.}
\lref\rAg{S. Agmon, {\it Lectures on Exponential Decay of Solutions of
Second-Order Elliptic Equations}, Princeton University Press (Princeton) 1982.}
\lref\rY{P. Yi, hep-th/9704098.}
\lref\rDLhet{ D. Lowe, hep-th/9704041.}
\lref\rqm{R. Flume, Ann. Phys. {\bf 164} (1985) 189\semi
M. Baake, P. Reinecke and V. Rittenberg, J. Math. Phys. {\bf 26} (1985) 1070.}
\lref\rbb{K. Becker and M. Becker, hep-th/9705091, \np{506}{1997}{48}\semi
K. Becker, M. Becker, J. Polchinski and A. Tseytlin, hep-th/9706072,
\prd{56}{1997}{3174}.}
\lref\rpw{J. Plefka and A. Waldron, hep-th/9710104, \np{512}{1998}{460}.}
\lref\rhs{M. Halpern and C. Schwartz, hep-th/9712133, Int. J. Mod. Phys. {\bf A13} (1998)
4367.}
\lref\rfrolich{J. Frohlich, G. M. Graf, D. Hasler, J. Hoppe and S.-T. Yau, 
hep-th/9904182. }
\lref\rlimit{N. Seiberg hep-th/9710009, \prl{79}{1997}{3577}\semi
A. Sen, hep-th/9709220.}
\lref\rentin{D.-E. Diaconescu and R. Entin, hep-th/9706059,
\prd{56}{1997}{8045}.}
\lref\rgreen{M. B. Green and M. Gutperle, hep-th/9701093, \np{498}{1997}{195}.}
\lref\rpioline{B. Pioline, hep-th/9804023.}
\lref\rgl{O. Ganor and L. Motl, hep-th/9803108.}
\lref\rds{M. Dine and N. Seiberg, hep-th/9705057, \pl{409}{1997}{209}.}
\lref\rberg{E. Bergshoeff, M. Rakowski and E. Sezgin, \pl{185}{1987}{371}.}
\lref\rBHP{M. Barrio, R. Helling and G. Polhemus, hep-th/9801189.}
\lref\rper{P. Berglund and D. Minic, hep-th/9708063, \pl{415}{1997}{122}.}
\lref\rspin{P. Kraus, hep-th/9709199, \pl{419}{1998}{73}\semi
J. Harvey, hep-th/9706039\semi
J. Morales, C. Scrucca and M. Serone, hep-th/9709063, \pl{417}{1998}{233}.}
\lref\rdine{M. Dine, R. Echols and J. Gray, hep-th/9805007; hep-th/9810021.}
\lref\rber{D. Berenstein and R. Corrado, hep-th/9702108, \pl{406}{1997}{37}.}
\lref\rnonpert{A. Sen, hep-th/9402032, \pl{329}{1994}{217}; hep-th/9402002,  
Int. J. Mod. Phys. {\bf
A9} (1994) 3707\semi
N. Seiberg and E. Witten, hep-th/9408099, \np{431}{1995}{484}.
}
\lref\rberkooz{M. Berkooz and M. R. Douglas, hep-th/9610236, 
\pl{395}{1997}{196}.}

\lref\rpss{S. Paban, S. Sethi and M. Stern, hep-th/9805018, \np{534}{1998}{137}.}
\lref\rpsst{S. Paban, S. Sethi and M. Stern, hep-th/9806028. }
\lref\rperiwal{V. Periwal and R. von Unge, hep-th/9801121.}
\lref\rfer{M. Fabbrichesi, G. Ferreti and R. Iengo, hep-th/9806018.}
\lref\rtoappear{S. Sethi and M. Stern, hep-th/0001189.}
\lref\rtr{A.~A.~Tseytlin, hep-th/9604035, \np{475}{1996}{149}. }
\lref\rnoncomm{A. Connes, M. R. Douglas and A. Schwarz, hep-th/9711162, JHEP 9802:003, 1998.}
\lref\rkraus{E. Keski-Vakkuri and P. Kraus, hep-th/9712013, \np{529}{1998}{246}.}
\lref\ralwis{S. P. de Alwis, hep-th/9806178, \prd{59}{1999}{044029}. }
\lref\rbkbk{K. Becker and M. Becker, hep-th/9712238, \prd{57}{1998}{6464}.}

\Title{\vbox{\hbox{hep-th/0002131}
\hbox{DUK-CGTP-00-04, IASSNS--HEP--99/117 }}}
{\vbox{\centerline{The Structure of the D0-D4 Bound State}}}

\centerline{Savdeep
Sethi$^\ast$\footnote{$^1$} {sethi@sns.ias.edu} and Mark
Stern$^\dagger$\footnote{$^2$} {stern@math.duke.edu} }

\medskip\centerline{$\ast$ \it School of Natural Sciences, Institute for
Advanced Study, Princeton, NJ 08540, USA}
\medskip\centerline{$\dagger$ \it Department of Mathematics, Duke University,  
Durham, NC 27706, USA}

\vskip 0.2in

We derive a set of equations for the wavefunction describing the marginal bound state
of a single D0-brane with a single D4-brane. These are equations determining the vacuum
of an $N=8$ abelian gauge theory with a charged hypermultiplet.
We then solve these equations for the most
general possible zero-energy solution using a Taylor series. We find that there are 
an infinite number of such solutions of which only one must be normalizable. 
We explore the
structure of a normalizable solution under the assumption of an asymptotic expansion. 
Even the leading terms in the asymptotic series, which should reflect the supergravity 
solution, are unusual. Through the
$Spin(5)$ flavor symmetry, the modes which are massive at long distance actually influence 
the leading behavior. Lastly, we show that the vacuum equations can quite remarkably
be reduced to a single equation involving one unknown function. 
The resulting equation has a surprisingly simple and suggestive form.

\vskip 0.1in
\Date{2/00}

%
\newsec{Introduction}

A single D0-brane and a single D4-brane form a marginal bound state \rhmon. The
low-energy dynamics of a D0-brane in the presence of a D4-brane is described
by a quantum mechanical Yang-Mills theory with eight supercharges. The theory 
has 
a $U(1)$ vector multiplet coupled to a charged hypermultiplet \rberkooz. 
With a single D4-brane,
there is only a Coulomb branch. The same 
quantum mechanics appears in the problem of counting H-monopole ground states
in the toroidally compactified heterotic string \rwitthmon. While the structure 
of
vacuum wavefunctions in marginally bound systems is typically very difficult to 
analyze, this particular theory has a number of simplifying features. The aim of 
this paper is to study the vacuum wavefunction of this $0+1$-dimensional gauge theory
with eight supercharges. 

Our goal is to gain insight into a number of issues. For example, how do we go about 
uncovering the structure of 
threshold wavefunctions? It is not even clear how to formulate reasonable questions about   
a system as complex as the quantum mechanics describing many D0-branes. 
Another major issue is how the full quantum mechanics resolves the singularity of 
the moduli 
space metric. The vector multiplet contains five scalars $x^\mu$. For large
$r=|x|$, the effective action describing the Coulomb branch dynamics 
should be a reasonable description
of the physics \rDKPS. The metric on the Coulomb branch is protected by 
supersymmetry
\rentin\
 and takes the form,
\eqn\metric{ ds^2 = \left( {1\over g^2}+ {1\over r^3}\right) (dx)^2, }
where $g^2$ is the Yang-Mills coupling constant. We can express $g^2$ in terms 
of 
the type IIA string scale $M_s$ and coupling constant $g_s$,
$$ g^2 = g_s  {M_s^3}. $$ 
This is the only scale in 
the theory. In the following sections, we set $g^2=1$ for simplicity.
The tube-like metric \metric\ has a singularity at $r=0$ which is  
resolved by the
full quantum mechanics.  Metrics with a similar structure appear
in D1-D5 systems.    

In the following section, we present the supercharges and describe the symmetries of the
problem. The flavor symmetry is $Spin(5) \times SU(2)_R$, and the unique vacuum is 
invariant under this symmetry \rtoappear. In section three, we derive the general
form of a gauge invariant and flavor invariant wavefunction. A general wavefunction
depends on $11$ functions of two variables, $r$ and $y$. As above, $r$ is a radial
coordinate for the $5$ scalars $x^\mu$ of the vector multiplet. The hypermultiplet has 
$4$ scalars $q_i$, which parametrize the massive directions. We take $y = |q|$. We then
derive a set of differential equations that any zero energy wavefunction must obey. 

In section four, we analyze the implications of these differential equations. We can
immediately reduce the number of unknown functions from $11$ to $7$. These $7$ functions 
satisfy
$14$ first order coupled partial differential equations in $2$ variables. We point out some
intriguing features of these equations: in particular, there is an interesting formal
method of reducing the number of functions and equations in which the harmonic 
oscillator plays a key role. 
This method for collapsing the system of equations 
has similarities to the technique used to find non-renormalization theorems \rpss.  
We, however,  
proceed in a different direction. We solve
the $14$ equations exactly by Taylor expanding in the $y$ variable. This is an expansion
in the massive directions. The first interesting point is that the differential equations
alone do not determine the solution. There are actually an infinite number of zero
energy solutions. The `gauge' degrees of freedom are not a finite set of parameters 
as we might have expected, but an entire function of degrees of freedom.

Essentially, from the perspective of the Taylor series, the problem boils down 
to solving $2$
ordinary differential equations in $3$ unknown functions. All the other terms in the 
wavefunction
are determined by these $3$ functions of $r$. The condition that must uniquely specify the
solution is normalizability. Rather remarkably, this global condition is strong enough 
to fix
an entire arbitrary function. This seems to hint that the kind of principle that should 
underly M theory involves global rather than local constraints. In a vague sense, this 
is reminiscent of holography. 

In section five, we turn to the practical problem of determining the normalizable solution. 
It turns out to be difficult to implement the global constraint of normalizability in any
nice way.  Instead, we
 expand the solution in an asymptotic series. This is akin
to solving the M theory equations of motion for the geometry of an M5-brane in a 
derivative expansion. We find that even the leading terms in the solution are unusual. These
terms should match a supergravity analysis. However, the structure of these terms is 
strongly
dictated by invariance under the $Spin(5)$ flavor symmetry. Invariance under $Spin(5)$ is 
not a statement about long distance, moduli space physics. It is a statement that 
requires knowledge of
both long and short distance physics because the $Spin(5)$ generators act on both the massless
and massive degrees of freedom. Somewhat contrary to intuition, we find that the 
vacuum state for the massive degrees of freedom at large $r$ is a sum
of $3$ representations of $Spin(5)$: a spherically symmetric $\1$, a $\5$ and a ${\bf 14}$.
 
The leading terms in the bound state wavefunction $\Psi$ have the form,
\eqn\leading{ \Psi \sim  {1\over r^3} e^{-{r y^2\over 2}} |b_1> + {x^\mu \over r^4}
e^{- {r y^2\over 2}} |b_2>^\mu + {x^\mu x^\nu
\over r^5} e^{-{r y^2\over 2}} |b_3>^{\mu \nu}, }
where the $|b_i>$ are constructed from fermions. It seems unlikely that this asymptotic
form could have been determined from low-energy considerations alone. In this sense, the
massive degrees of freedom, through the $Spin(5)$ flavor symmetry, are important even at
arbitrarily long distances. The structure of $\Psi$ in \leading\ really begs for  
an interpretation
 both in terms of the supergravity solution of the D0-D4 brane \rtr, and in terms of 
the DLCQ description of an M5-brane \rberkooz\  via Matrix theory \rBFSS. 

We proceed to compute the general form of $\Psi$ in an asymptotic expansion. This takes 
us well beyond supergravity. The corrections to the
leading terms \leading\ take the form of a perturbation series in the coupling constant $g^2$. 
Each correction depends on an a priori unknown constant, and we give a prescription for 
determining
this constant. This amounts to summing up the higher derivative
corrections to the supergravity solution for an M5-brane. Is the asymptotic solution 
actually
convergent as $r\r 0$? We know of no non-perturbative effects in the abelian gauge theory 
that could be relevant at
short distances. However, this does not prove that the solution is convergent. It would be
interesting to sum up a sufficient number of terms in the asymptotic series to see whether
the solution is well-behaved as $r$ becomes small. This would clarify how the singularity in
the metric is resolved from the perspective of a derivative expansion. 

We note that finding the bound state
wavefunction in an asymptotic series is much like trying to understand M theory in 
a derivative expansion. In section six, we present a quite different reduction of 
our initial $14$ vacuum equations, one that
perhaps an M theorist might use. The result is quite incredible. The entire problem
reduces to solving a {\it scalar} equation of the form,
\eqn\surprise{ \left( \Delta + \vec{B} \cdot \nabla + W \right) u =0, }
where $\Delta = \partial_r^2 + \partial_y^2$ and $u$ is a particular combination of the 
functions that appear in the bound state wavefunction. The vector field $ \vec{B}$ and
potential $W$ are rational functions of $r$ and $y$. Equation \surprise\ is both
simple and highly suggestive. It would be very interesting to find the solution
to \surprise\ either analytically or numerically. Supergravity and the structure
of the derivative expansion should emerge from the asymptotics of the resulting
solution for the bound state wavefunction. It would also be extremely interesting
to generalize this reduction to non-abelian gauge theories. This might possibly
help us understand how to define M theory through Matrix theory \rBFSS. It does not 
seem too unlikely to us that the ability to `deprolong' the initial vacuum equations
to get \surprise, as described in section six, is tied to supersymmetry and invariance
theorems \rtoappear. 

There are many additional directions to
explore. Some of the simplifying
features of the D0-D4 system remain when we add more hypermultiplets. However, 
there will
now be a Higgs branch and the zero energy wavefunctions will spread in an 
interesting
way onto the Higgs branch. Turning on $B$-fields makes the gauge theory on the D4-brane
non-commutative \rnoncomm. Certain choices of $B$-field should change the asymptotic 
behavior from polynomial decay to exponential decay. It would also be interesting to 
actually match the asymptotic structure of the bound state wavefunction with  higher derivative
corrections to the supergravity solution, like those generated by the $R^4$ terms 
\refs{\rkraus, \ralwis, \rbkbk}.  

\newsec{The D0-D4 Quantum Mechanics}
\subsec{The vector multiplet supercharge}

The D0-D4 system is obtained by dimensionally reducing $N=1$ abelian Yang-Mills 
with a
single charged hypermultiplet from six dimensions. The symmetry group consists 
of the
R-symmetries\foot{The symmetry group,
including both gauge and flavor symmetries, is not globally a product. There are
discrete identifications. However, for this analysis we only need the Lie 
algebra 
generators so we can ignore global identifications.}  
$Spin(5)\times SU(2)_R \sim Sp(2)\times Sp(1)_R$. The Hamiltonian
is invariant under the symmetry group while the eight real supercharges 
transform in the $( \4, \2)$ representation.

Let us begin with the vector
multiplet which contains the five scalars $x^\mu$ transforming in the 
$({\bf 5}, {\bf 1})$ of the symmetry group. Let $p^\mu$ be the associated 
canonical momenta obeying,
\eqn\bosquant{ \com{x^\mu}{p^\nu} = i\delta^{\mu\nu}. }
Associated to these bosons are eight real fermions $\l_a$ where $a=1,\ldots,8$  
transforming in the $( \4, \2)$ representation of the symmetry group.
These fermions obey the usual quantization relation, 
\eqn\fermquant{ \acom{\l_{a}}{\l_{b}} = \delta_{ab}.} 
Let $\g^\mu$ be hermitian real 
gamma matrices which obey, 
\eqn\gmat{ \acom{\g^\mu}{\g^\nu} = 2 \delta^{\mu\nu}. }
An explicit basis for these gamma matrices along with a 
discussion of the symmetry group action is given in Appendix A. 

To write the vector multiplet supercharge, we introduce an auxiliary field $D$
which transforms as $({\bf 1},{\bf 3})$ under the symmetry group. The $D$-term is
independent of $x^\mu$. The vector 
multiplet supercharge is given by:
\eqn\vectorsuper{Q_{a}^v = \left( \g^\mu p^\mu \l \right)_a 
+ D_{ab} \l_b.  }
The real anti-symmetric matrix $D$ commutes with $\g^\mu$ because
the $Sp(1)_R$ and $Sp(2)$ actions commute. The $D$-term must also satisfy,
\eqn\Dconst{ D_{ac} D_{bc} = - \delta_{ab}\, |D|^2.}
It is then not hard to check that:
\eqn\vectoralg{ \acom{Q_{a}^v}{Q_{b}^v} = \delta_{ab} \left\{ p^2 + |D|^2 
\right\}.}
Under a symmetry transformation $(U, g) \in Sp(2)\times Sp(1)_R$, we note that
\eqn\symm{\eqalign{ \g^\mu p^\mu \, \r \,  U \g^\mu p^\mu U^{-1}, \qquad 
\l \, \r \,  U g \l, \qquad 
D \, \r \, g D g^{-1}, }}
so that
\eqn\symmq{Q^v \, \r \, U g Q^v. }

\subsec{The hypermultiplet supercharge}

A hypermultiplet contains four real scalars which we can package into a 
quaternion
$q$ with components $q^i$ where $i=1,2,3,4$. This field transforms 
as $({\bf 1}, {\bf 2})$ under the symmetry group. We again introduce canonical 
momenta
$p_i$ satisfying the usual commutation relations.

The hypermultiplet is charged under the $U(1)$ gauge symmetry so we need to 
determine
how $U(1)$ acts on $q$. The $q^i$ parametrize $\IR^4$ so the symmetry group 
acting on the
hypermultiplet must sit inside, 
$$SO(4) \sim Sp(1)_L \times Sp(1)_R.$$ 
Gauge transformations and $Sp(1)_R$ transformations commute. Therefore, the 
$U(1)$
gauge symmetry sits inside $Sp(1)_L$. We choose to generate the 
gauge symmetry by left 
multiplication on $q$ by ${ I}$. The hermitian generator of gauge 
transformations on the bosons is then given by, 
\eqn\gaugeboson{ G_b = W_{12} + W_{34},}
where 
\eqn\defW{ W_{ij}  = q_i p_j - q_j p_i.}  
The superpartner to $q$ is a real fermion $\f_a$ with $a=1,\ldots, 8$ 
satisfying,
\eqn\secondquant{ \acom{\f_{a}}{\f_{b}} = \delta_{a b},}
and transforming in the $({\bf 4}, {\bf 1})$ representation. In terms of the 
$s^j$
operators given in Appendix A, the free hypermultiplet charge takes the form
\eqn\freehyper{  Q^{h_f}_a = s^j_{ab} \f_b \, p_j.} 
Note that since the $s^j$ implement right multiplication by a quaternion, they 
commute with $\g^\mu$. This free charge obeys the algebra, 
$$ \acom{Q_{a}^{h_f}}{Q_{b}^{h_f}} = \delta_{ab} \, p_i p_i.$$
Invariance of \freehyper\ under the $U(1)$ gauge symmetry requires that
\eqn\gaugefermion{ G_f =  - {i\over 2}\, s^2_{ab} \f_a \f_b}
generate gauge transformations on $\f$. The total generator of the $U(1)$ gauge
symmetry is then given by, 
\eqn\gauge{ G = G_b + G_f = W_{12} + W_{34} - {i\over 2} \f s^2 \f.  }
The full hypermultiplet
supercharge $Q^h$ also includes couplings to the vector multiplet,
\eqn\hypercharge{Q^h_a =  s^j_{ab} \f_b \, p_j + (\g^\mu s^j s^2)_{ab} \, 
\f_b \, x^\mu q_j.}
The form of the interaction term in \hypercharge\ is fixed up to an overall
constant by symmetry. The $s^2$ appearing in the interaction term is needed to 
ensure
that $Q^h$ is gauge-invariant.
The charge obeys the algebra:
\eqn\hyperalg{ 
\acom{Q^h_{a}}{Q^h_{b}} = \delta_{ab} \left\{ p_i p_i + |x|^2 |q|^2 - {i\over 
4}\,
x^\mu \f \g^\mu s^2 \f  \right\} + 2 \g^\mu_{ab} x^\mu G. }
As we expect, the supersymmetry algebra only closes on the Hamiltonian up to 
gauge transformations.

\subsec{The coupled system}

The full supercharge $Q$ is the given by,
\eqn\fullcharge{Q = Q^v + Q^h,}
where we define the $D$-term in the following way: using the components $q^i$
and matrices $s^i$, we can define a quaternion and its conjugate 
which act by right multiplication on $\l$:
$$ \eqalign{ q^R &  = s^1 q^1 + s^2 q^2 + s^3 q^3 + s^4 q^4 \cr
\bar{q}^R & = s^1 q^1 - s^2 q^2 - s^3 q^3 - s^4 q^4.} $$
We can then write the $D$-term in the form, 
\eqn\Dterm{ D_{ab} = {1\over 2} \, \left( q^R \, s^2 \, \bar{q}^R \right)_{ab}.}
This $D$-term obeys \Dconst, 
$$ \eqalign{ \left( D^2 \right)_{ab} & = \delta_{ab} \, |D|^2 \cr
& = \delta_{ab} \, {1\over 4} \left( q \bar{q} \right)^2.} $$ 
The full charge obeys the algebra:
\eqn\completesusy{\eqalign{ \acom{Q_{a}}{Q_{b}}& = \delta_{ab} 
\left\{ p^\mu p^\mu + |D|^2 + p_i p_i + |x|^2 |q|^2 + \ldots  \right\} 
+ 2 \g^\mu_{ab} x^\mu G \cr
& = \delta_{ab}\, 2 H +  2 \g^\mu_{ab} x^\mu G. }}
The omitted terms are bilinears in the fermions whose exact form we will not
need. The bosonic potential $V$ appearing in \completesusy\ is given by,
\eqn\V{ V = |x|^2 |q|^2 + {1\over 4} \, |q|^4. }
Since we have coupled a single hypermultiplet to the $U(1)$ vector multiplet, 
the only flat direction is $q=0$ and there is no Higgs branch. 

\newsec{Deriving Equations for the Vacuum Wavefunction}

To be consistent with predictions from string duality, there should be a unique
vacuum wavefunction for this quantum mechanical gauge theory \rwitthmon. 
An index argument proves that there is at least one normalizable vacuum
wavefunction \rhmon. Coupled with a recent invariance theorem \rtoappear, the 
index result implies that the ground state is unique. 

On quantization, the fermions $\l$ and $\f$ act as gamma matrices on a 
$256$-dimensional
spinor wavefunction. A priori, the vacuum wavefunction then consists of 
$256$ complex functions of the $9$ bosonic variables $x^\mu$ and $q^i$. However, 
we can
significantly simplify the problem by using symmetries. First note that any 
state
in the Hilbert space $|s>$ must be gauge-invariant, 
\eqn\gaugeinv{ G |s> =0.}
Further, all states can be grouped into representations of the global 
$Sp(2)\times Sp(1)_R$ symmetry group. The $Sp(2)$ is generated by the operators,
\eqn\spingen{ T^{\mu\nu} = X^{\mu\nu} -{i\over 4} \, \g^{\mu\nu}_{ab}\left( 
\l_a\l_b + 
\f_a\f_b\right),}
where
\eqn\defX{X^{\mu\nu} = x^\mu p^\nu - x^\nu p^\mu.} 
The three generators of $Sp(1)_R$ correspond to right multiplication by $I,J,K$
and in accord with prior notation, we will denote them by $\ss^i$:
\eqn\defss{\eqalign{ \ss^2 & = W_{12} - W_{34} + {i\over 2}\, \l s^2 \l \cr
\ss^3 & = W_{13} + W_{24} + {i\over 2}\, \l s^3 \l \cr
\ss^4 & = W_{14} - W_{23} + {i\over 2}  \, \l s^4 \l. }}
The unique ground state $\Psi$ must be invariant under the actions of  $Q_a,
G, T^{\mu\nu}$ 
and $ \ss^i$,
$$ Q_a \Psi = G \Psi = T^{\mu\nu} \Psi = \ss^i \Psi =0.$$
These constraints are quite powerful; for example, they 
allow us to replace a differential operator $X^{\mu\nu}$ 
by an algebraic one ${i\over 4} \, \g^{\mu\nu}_{ab}\left( \l_a\l_b + 
\f_a\f_b\right).$
There are multiple ways to derive equations for the vacuum wavefunction. We will 
describe two 
approaches which we used to derive these equations.\foot{Using two different 
approaches 
helped enormously in the search for errors.}

\subsec{Radial coordinates}

In the first approach, 
we can rewrite the supercharge \fullcharge\ in terms of radial coordinates:
$$ r^2 = |x|^2, \qquad y^2 = |q|^2. $$
The charge takes the form, 
\eqn\radialcharge{ \eqalign{ Q_a &= (\g^\mu \l)_a {x^\mu\over r} p_r + 
(s^j \f)_a {q^j\over y} p_y + (\g^\mu \l)_a {x^\nu\over r^2} X^{\nu\mu}
+ (s^j \f)_a {q^k\over y^2} W_{kj} \cr
&  + (\g^\mu s^j s^2 \f)_a \, x^\mu q_j +(D\l)_a \cr
& =(\g^\mu \l)_a {x^\mu\over r} p_r + 
(s^j \f)_a {q^j\over y} p_y + M_a.}}
We have lumped the angular derivatives and non-derivative terms into the 
operator
$M_a$. What is particularly nice about $M_a$ is that we can replace all the 
derivative operators by bilinears in fermions. As we noted before, $Sp(2)$
invariance allows us to replace $X^{\nu\mu}$ by
$${i\over 4} \, \g^{\nu\mu}_{ab}\left( \l_a\l_b + \f_a\f_b\right).$$
However, we can also replace $ q^k W_{kj} $ by a bilinear in fermions using 
$Sp(1)_R$ invariance. Using \defss, we note that:
\eqn\replace{\eqalign{ q^k W_{k1} &= {i\over 2} \left\{ q^2 \l s^2 \l +
q^3 \l s^3 \l + q^4 \l s^4 \l  \right\} \cr
q^k W_{k2} &=  {i\over 2}\left\{q^4 \l s^3 \l - q^1 \l s^2 \l 
 - q^3 \l s^4 \l  \right\} \cr
q^k W_{k3} &= {i\over 2} \left\{ q^2 \l s^4 \l -
q^1 \l s^3 \l - q^4 \l s^2 \l  \right\} \cr 
q^k W_{k4} &=  {i\over 2}\left\{ q^3 \l s^2 \l -
q^1 \l s^4 \l - q^2 \l s^3 \l  \right\}. }}
Therefore, $M_a$ is a completely algebraic operator. 

Using the $Sp(2)$ symmetry, we can then 
rotate $x$ to the special point where $x^1 \neq 0$ and $x^\mu=0$ for  $\mu>1$. 
Likewise,
we can rotate $q$ using $Sp(1)_R$ to the point $q^1\neq 0$ and $q^i=0$ for 
$i>1$. At this point,
$r=|x^1|$ and $y=|q^1|$. Since all
angular derivatives in $Q_a$ are replaced by algebraic operators, there is no 
difficulty in
restricting $\Psi$
to this point. The question of determining
$\Psi$ at this point then reduces to finding coupled differential equations in 
two variables. The
form of $\Psi$ at an arbitrary choice of $x$ and $q$ can then be obtained by 
applying the 
rotation generators 
\spingen\ and \defss.

\subsec{Symmetries and the fermion Hilbert space}

In the second approach which we will use for the rest of the paper, we will 
first solve the 
invariance conditions explicitly for the most general 
possible invariant wavefunction. As we will show, the most general wavefunction 
depends 
on $11$ functions of $r$ and $y$. 
We will then derive coupled equations for these functions from the requirement 
that the 
wavefunction have zero energy.

The first step is to construct the fermion Hilbert space. We need to complexify 
our real fermions 
and build a Fock space: 
\eqn\complexify{ \eqalign{ \sqrt{2}\, du_1  = \l_1 + i \l_{2} \qquad\quad &  
\sqrt{2}\, dv_1
  = \f_1 + i \f_{2}, \cr
                           \sqrt{2}\, du_2  = \l_3 - i \l_{4} \qquad\quad &  
\sqrt{2}\, dv_2
  = \f_3 - i \f_{4},\cr
                           \sqrt{2}\, du_3  = \l_5 + i \l_{6} \qquad\quad &  
\sqrt{2}\, dv_3
  = \f_5 + i \f_{6},\cr
                           \sqrt{2}\, du_4  = \l_7 - i \l_{8} \qquad\quad &  
\sqrt{2}\, 
dv_4 = \f_7 - i \f_{8}.\cr}} 
It is natural to think
of $du_a$ and $dv_a$ as one-forms obeying the relation,
$$ \acom{du_a}{du_b^*} = \delta_{ab} \qquad \acom{dv_a}{dv_b^*} = \delta_{ab}, 
$$
where $a=1,\ldots,4$. 
Wavefunctions in the Hilbert space are then $(p,q)$ 
forms where $p$ and $q$ are the $du_a$ and $dv_a$ degrees, respectively. We 
choose the Fock
vacuum or $(0,0)$ form to satisfy,
$$ du_a^* |0> = dv_a^* |0> =0. $$
Note that the complex conjugate of a $(p,q)$-form is a $(4-p, 4-q)$-form. 
If the ground state is unique then it is bosonic so the form degree must be 
even. 

With our choice of complexification \complexify, the $Sp(2)$ generators 
$T^{\mu\nu}$ acting
on forms preserve degree. Actually, the generators preserve $p$ and $q$ 
separately so a $(p,q)$
form is mapped to a $(p,q)$ form. The $Sp(2)$ generators naturally split into 
commuting generators 
for an
$Sp(2)_b$ acting on bosons and an $Sp(2)_f$ acting on fermions. In turn, the 
$Sp(2)_f$ splits 
into an $Sp(2)_{f_p}$ acting on $du$ with generators,
$$-{i\over 4} \, \sum_{a,b} \g^{\nu\mu}_{ab}\, \l_a\l_b, $$
and an $Sp(2)_{f_q}$ acting on $dv$ with generators,
$$-{i\over 4} \, \sum_{a,b} \g^{\nu\mu}_{ab} \f_a\f_b.$$ 
We can now employ some 
group theory to see how the various $128$ bosonic forms transform under
$Sp(2)_f$. Let us start with the $(p,0)$ forms which appear in the following 
representations:
$$
 \vbox{\offinterlineskip
\hrule
\halign{&\vrule#&
  \strut\quad\hfil#\quad\cr
height2pt&\omit&&\omit&\cr
& $p$ \hfil&& $Sp(2)_f$ rep. &\cr
height2pt&\omit&&\omit&\cr
\noalign{\hrule}
height2pt&\omit&&\omit&\cr
& 0 && \hfil $\1$\hfil&\cr
& 1 &&  \hfil $\4$\hfil&\cr
& 2 && $\,\,\,$ \hfil  ${\5\oplus\1}$\hfil&\cr
& 3 && \hfil $\4$\hfil&\cr
& 4 && \hfil $\1$\hfil&\cr
height2pt&\omit&&\omit&\cr}
\hrule}$$
We wedge the (odd) even $(p,0)$ forms with the (odd) even $(0,q)$ forms to get 
the $128$
bosonic forms. The following representations appear from wedging even forms with 
even forms,
$$ (\1)^{10} \oplus (\5)^6 \oplus {\bf 10}\oplus {\bf 14},$$
while from wedging odd forms with odd forms, we find:
$$ (\1)^{4} \oplus (\5)^4 \oplus ({\bf 10})^4. $$ 
We can immediately discard forms transforming in the $ {\bf 10}$ representation. 
A tensor
say $a^{\mu\nu}$ transforming in the $ {\bf 10}$ is antisymmetric in $\mu,\nu$ 
so contraction with $ x^\mu x^\nu $ to get a singlet of the full $Sp(2)$ gives 
zero.

Let us now constrain our Hilbert space further by imposing invariance
under $Sp(1)_R$ and the gauge symmetry. We can rewrite the generators \defss\ in 
terms of our complex fermions:
\eqn\newdefss{\eqalign{ \ss^2 & = W_{12} - W_{34} +  \sum_a du_a du_a^* - 2 \cr
\ss^3 & = W_{13} + W_{24} - i \left( du_1 du_2 + du_1^* du_2^* + du_3 du_4 
+ du_3^* du_4^* \right) \cr
\ss^4 & = W_{14} - W_{23} +  \left( du_1 du_2 - du_1^* du_2^* + du_3 du_4 
- du_3^* du_4^* \right). }}
Likewise for the gauge symmetry, 
\eqn\newgauge{G = W_{12} + W_{34} -   \sum_a dv_a dv_a^* + 2.}
Note that the operator $W_{ij}$ has eigenvalues $n$ and $-n$ with corresponding 
eigenfunctions,
$$ z_{ij}^n = (q_i + i q_j)^n, \qquad \bar{z}_{ij}^n= (q_i - i q_j)^n. $$
What does invariance under \newdefss\ and \newgauge\ imply? By taking the sum 
and difference
of $G$ and $\ss^2$, we see that we should restrict to $(p,q)$ forms
\eqn\gaugeres{ (z_{12})^{q-p\over 2} (z_{34})^{{(p+q)\over 2}-2} |p,q>, }
which can be multiplied by a function of both $|z_{12}|^2$ and $|z_{34}|^2$. 
The remaining two generators $\ss^3$ and $\ss^4$ change the value of $p$.  
It is natural to study the complex combinations $\ss^3-i\ss^4$ and 
$\ss^3+i\ss^4$, 
which raise and lower the value of $p$:
\eqn\remainingtwo{\eqalign{ s^+ = \ss^3 - i\ss^4 & = z_{12} p_{34} - 
\bar{z}_{34}\bar{p}_{12}
 - 2 i \left( du_1 du_2 + du_3 du_4 \right) \cr
 s^- = \ss^3 + i\ss^4 & = \bar{z}_{12}\bar{p}_{34} - z_{34} p_{12}
 - 2 i \left( du_1^* du_2^* + du_3^* du_4^* \right), }}
where $p_{ij} = p_i - i p_j$ and $\bar{p}_{ij} = p_i + i p_j$.

Again the generators of $Sp(1)_R$ split into an $Sp(1)_b$ acting on bosons and 
an 
$Sp(1)_f$ acting on fermions. It is easy to see how the $(p,q)$-forms fall into 
representations
of $Sp(1)_f$. The three singlets under $Sp(2)_{f_p}$ denoted $|0,q>, 
|2,q>_{\1_p}, |4,q>$
transform in the $\3$ of $Sp(1)_f$. For the choice $q=0,4$, 
we can construct 
one singlet under the full $Sp(1)_R$ which can be multiplied by an arbitrary 
function of $y$. For the
case $q=2$, we can construct two singlets under $Sp(1)_R$ by tensoring with 
either the $\1$
or the $\5$ of  $Sp(2)_{f_q}$. Let us denote the $\5$ of  $Sp(2)_{f_q}$ by 
$|2>_{\5_q}^\mu$. The 
explicit $Sp(1)_R$ singlets are then given by the forms, 
$$ \eqalign{ & \left\{ \bar{z}_{34}^2 + \bar{z}_{12} \bar{z}_{34} (du_1 du_2 + 
du_3 du_4)
+ \bar{z}_{12}^2 du_1 du_2 du_3 du_4 \right\} |0>, \cr
&  \big\{ z_{12} \bar{z}_{34} + {1\over 2} (|z_{12}|^2 - |z_{34}|^2) 
(du_1 du_2 + du_3 du_4)
- \bar{z}_{12} z_{34} du_1 du_2 du_3 du_4 \big\} 
\times \cr & (dv_1 dv_2 + dv_3 dv_4) |0>, \cr
&  \big\{ z_{12} \bar{z}_{34} + {1\over 2} (|z_{12}|^2 - |z_{34}|^2) 
(du_1 du_2 + du_3 du_4)
- \bar{z}_{12} z_{34} du_1 du_2 du_3 du_4 \big\} |2>_{\5_q}^\mu, \cr
 & \left\{ z_{12}^2 - z_{12} z_{34} (du_1 du_2 + du_3 du_4)
+ z_{34}^2 du_1 du_2 du_3 du_4 \right\} dv_1 dv_2 dv_3 dv_4 |0>. }$$ 

The $\5$ of $Sp(2)_{f_p}$ denoted $|2>^\mu_{\5_p}$ decomposes into five singlets 
under
 $Sp(1)_f$. The form  $|2>_{\5_p}$ can therefore only appear with a function of 
$y$. 
To satisfy the constraint \gaugeres, we must then tensor $|2>^\mu_{\5_p}$ with a 
$q=2$ form 
constructed from $dv$. The two choices are either the  $\1$
or the $\5$ of  $Sp(2)_{f_q}$. This gives three additional
possibilities denoted, 
$$ |2,2>_{\1}, \quad |2,2>^\mu_{\5}, \quad  |2,2>^{\mu\nu}_{\bf 14},$$
where the subscript denotes the representation under the full $Sp(2)_f$. The 
construction of
these forms is described in Appendix B. 

Lastly, we need to consider
the case of odd $p$. The $|1>_{\4_p}$ and  $|3>_{\4_p}$ forms combine to form a 
doublet
under $Sp(1)_f$. By tensoring with either the  $|1>_{\4_q}$ or the $|3>_{\4_q}$ 
forms, 
we can construct the following four $Sp(1)_R$ invariants:
$$ \eqalign{ 
& \left\{ z_{12} - z_{34} (du_1 du_2 + du_3 du_4) \right\} |1,3>_{\1} \cr 
& \left\{ z_{12} - z_{34} (du_1 du_2 + du_3 du_4) \right\} |1,3>_{\5}^\mu \cr 
& \left\{\bar{z}_{34} + \bar{z}_{12} (du_1 du_2 + du_3 du_4) \right\} |1,1>_{\1} 
\cr 
& \left\{\bar{z}_{34} + \bar{z}_{12} (du_1 du_2 + du_3 du_4) \right\} 
|1,1>_{\5}^\mu. \cr }
$$
Again the subscript denotes the representation under the full $Sp(2)_f$.
After imposing all the invariance constraints, we are therefore left with 
$11$ complex functions $f_i = f_i(r,y)$ appearing in the following way:
\eqn\listi{
f_1 |0,0> = f_1 \left\{ \bar{z}_{34}^2 + \bar{z}_{12} \bar{z}_{34} (du_1 du_2 + 
du_3 du_4)
+ \bar{z}_{12}^2 du_1 du_2 du_3 du_4 \right\} |0> }
\eqn\listii{ \eqalign{
 f_2 |0,4> = & f_2  \left\{ z_{12}^2 - z_{12} z_{34} (du_1 du_2 + du_3 du_4)  
+ z_{34}^2 du_1 du_2 du_3 du_4 \right\} \cr & \times dv_1 dv_2 dv_3 dv_4 |0> }}
\eqn\listiii{
  f_3 |2,2>_{\1}, \qquad   
  f_4 x^\mu |2,2>^\mu_{\5}, \qquad  
  f_5 x^\mu x^\nu  |2,2>^{\mu\nu}_{\bf 14},}
\eqn\listiv{ \eqalign{
 & f_6 |1,3> = f_6  \left\{ z_{12} - z_{34} (du_1 du_2 + du_3 du_4) \right\} 
|1,3>_{\1} \cr
 & f_7 |1,1> = f_7  \left\{\bar{z}_{34} + \bar{z}_{12} (du_1 du_2 + du_3 du_4) 
\right\} 
|1,1>_{\1} \cr}}
\eqn\listv{ \eqalign{
 & f_8 x^\mu |1,3>^\mu = f_8  x^\mu\left\{ z_{12} - z_{34} (du_1 du_2 + du_3 
du_4) 
   \right\} |1,3>_{\5}^\mu \cr
 & f_9 x^\mu |1,1>^\mu = f_9  x^\mu\left\{\bar{z}_{34} + \bar{z}_{12} (du_1 du_2 
+ 
du_3 du_4) \right\} |1,1>_{\5}^\mu}} 
\eqn\listx{\eqalign{ f_{10} |0,2> = & f_{10}
 \big\{ z_{12} \bar{z}_{34} + {1\over 2} (|z_{12}|^2 - |z_{34}|^2) 
(du_1 du_2 + du_3 du_4) \cr &
- \bar{z}_{12} z_{34} du_1 du_2 du_3 du_4 \big\} (dv_1 dv_2 + dv_3 dv_4) |0> }}
\eqn\listxi{\eqalign{ f_{11} x^\mu |0,2>^\mu = & f_{11} x^\mu
 \big\{ z_{12} \bar{z}_{34} + {1\over 2} (|z_{12}|^2 - |z_{34}|^2) 
(du_1 du_2 + du_3 du_4) \cr &
- \bar{z}_{12} z_{34} du_1 du_2 du_3 du_4 \big\} |2>_{\5_q}^\mu. }}
Our choice of normalization in constructing these forms is described in Appendix 
C.
We take the ground state $\Psi$ to be the sum of these eleven forms.

\subsec{Dynamical constraints} 

What remains is to determine the consequences of the eight equations,
\eqn\vacuum{ Q_a \Psi = 0.}
First note that each term in $Q_a$ can be assigned a parity $(\pm, \pm)$ 
according to whether it changes the parity of the wavefunction in $(x,q)$ 
respectively.
For example, the term
\eqn\aterm{  (\g^\mu s^j s^2 \f)_a \, x^\mu q_j }
has parity $(-,-)$ since it is odd in $x$ and odd in $q$. Likewise, each term in 
$\Psi$
has a definite parity. We can therefore isolate all terms in $ Q_a \Psi$ with a 
definite
parity. It is also sufficient to restrict to the case $a=1$ because our 
wavefunction
$\Psi$ is $Sp(2)$ invariant, but an $Sp(2)$ transformation rotates us from one 
choice of 
charge to another. 

We can then ask: what combinations give terms with parity $(+,-)$? A quick 
check of $Q_a$ acting on the possible forms composing $\Psi$ gives the following 
equation,  
\eqn\pmeq{ \eqalign{ &
 (s^j \f)_a p_j \left\{ \,  f_1 |0,0> + f_{10}|0,2> + f_2 |0,4>+ f_3 |2,2>_{\1} 
+  f_5 x^\mu x^\nu  |2,2>^{\mu\nu}_{\bf 14} \, \right\} + \cr 
& (\g^\mu s^j s^2 \f)_a \, x^\mu q^j  \left\{ f_4 x^\rho |2,2>^\rho_{\5} 
+ f_{11} x^\rho |0,2>^\rho \right\} \cr & + 
(\g^\mu \l)_a p^\mu \left\{ f_8 x^\rho |1,3>^\rho + f_9 x^\rho |1,1>^\rho 
\right\} 
 + (D \l)_a \left\{ f_6 |1,3> + f_7 |1,1> \right\} = 0. \cr}}
The terms giving $(-,-)$ satisfy:
\eqn\mmeq{ \eqalign{ &
(\g^\mu s^j s^2 \f)_a \, x^\mu q^j \big\{ \,  f_1 |0,0> + f_{10}|0,2> +
 f_2 |0,4>+ f_3 |2,2>_{\1} 
+  \cr & f_5 x^\mu x^\nu  |2,2>^{\mu\nu}_{\bf 14} \, \big\}  +  
  (s^j \f)_a p_j \left\{ f_4 x^\rho |2,2>^\rho_{\5} 
+ f_{11} x^\rho |0,2>^\rho \right\} \cr & + 
(D \l)_a\left\{ f_8 x^\rho |1,3>^\rho + f_9 x^\rho |1,1>^\rho \right\}  
 +  (\g^\mu \l)_a p^\mu \left\{ f_6 |1,3> + f_7 |1,1> \right\} = 0. \cr}}
From $(-,+)$, we find:
\eqn\mpeq{ \eqalign{ &
(\g^\mu \l)_a p^\mu \left\{ \,  f_1 |0,0> +  f_{10}|0,2> + f_2 |0,4>+f_3 
|2,2>_{\1} 
+  f_5 x^\mu x^\nu  |2,2>^{\mu\nu}_{\bf 14} \, \right\} + \cr 
& (D \l)_a  \left\{ f_4 x^\rho |2,2>^\rho_{\5} 
+ f_{11} x^\rho |0,2>^\rho \right\}  + 
(s^j \f)_a p_j\left\{ f_8 x^\rho |1,3>^\rho + f_9 x^\rho |1,1>^\rho \right\} \cr 
& +  (\g^\mu s^j s^2 \f)_a \, x^\mu q^j \left\{ f_6 |1,3> + f_7 |1,1> \right\} = 
0. \cr}}
The last equation follows from considering the $(+,+)$ terms, 
\eqn\ppeq{ \eqalign{ &
(D \l)_a \big\{ \,  f_1 |0,0> + f_{10}|0,2> + f_2 |0,4>+ f_3 |2,2>_{\1} 
+  f_5 x^\mu x^\nu  |2,2>^{\mu\nu}_{\bf 14} \, \big\} + \cr 
&  (\g^\mu \l)_a p^\mu \big\{ f_4 x^\rho |2,2>^\rho_{\5} 
+ f_{11} x^\rho |0,2>^\rho \big\} + \cr &
(\g^\mu s^j s^2 \f)_a \, x^\mu q^j \big\{ f_8 x^\rho |1,3>^\rho + f_9 x^\rho 
|1,1>^\rho \big\} 
+  \cr & (s^j \f)_a p_j \left\{ f_6 |1,3> + f_7 |1,1> \right\} = 0. \cr}}
In each equation, we set $a=1$ as a first simplification. After evaluating 
angular
derivatives,  we are free to rotate $q$ using $Sp(1)_R$ so that $q^1 \neq 0$
and $q^i=0$ for $i>1$. In a similar way, we  can consider the point $ x^1 \neq 
0$
with $x^\mu =0$ for $\mu>1$ after evaluating the $x$ angular derivatives.
With this choice of coordinates, $y=|q^1|$ and $r=|x^1|$. 

The first set of equations relate $f_1, f_7, f_9, f_{10}$ and $f_{11}$. These 
follow from 
considering
the $(4,1)$ forms and the $(3,0)$ forms in \pmeq, \mmeq\ and \mpeq, \ppeq\ 
respectively:
\eqn\relone{ \eqalign{ & r {\partial f_9 \over \partial r}  - 
y {\partial f_1 \over \partial y} - {y^2\over 2} f_7 + 5 f_9 -4 f_1 + 2 f_{10} 
=0, \cr
&  {\partial f_7 \over \partial r} +  r y^2 \left\{f_1 - {f_9\over 2} \right\} + 
2 r f_{11} = 0, \cr
& r {\partial f_9 \over \partial y} + y {\partial f_1 \over \partial r} + r y f_7 
=0, \cr
& {\partial f_7 \over \partial y} + r^2 y f_9 + {y^3\over 2} f_1  =0.
}}
Likewise, by considering the $(0,3)$ forms in \pmeq, \mmeq\ and the $(1,4)$ 
forms
in \mpeq, \ppeq, we find the following equations relating $f_2, f_6, f_8, 
f_{10}$ 
and $f_{11}$:
\eqn\reltwo{\eqalign{
&  r {\partial f_8 \over \partial r} + y {\partial f_2 \over \partial y} +
{y^2\over 2} f_6 + 5 f_8 + 4 f_2 - 2 f_{10}  =0, \cr
& {\partial f_6 \over \partial r} + r y^2 \left\{ f_2 + {f_8\over 2} \right\} + 2 
r f_{11}
=0, \cr
& y {\partial f_2 \over \partial r} - r {\partial f_8 \over \partial y} + r y f_6 
 =0, \cr
& {\partial f_6 \over \partial y} - r^2 y f_8 + {y^3\over 2} f_2  =0.}}
Note that equations \reltwo\ are the same as \relone\ under the identification:
\eqn\ident{ f_2 \r f_1 \qquad f_6 \r f_7 \qquad f_8 \r - f_9.}

It is easy to check there are no non-vanishing $(0,1),(4,3)$ and $ (1,0), (3,4)$ 
forms 
in \pmeq, \mmeq\ and \mpeq, \ppeq. 
This leaves equations coming from forms with degree $(2,1), (2,3)$ in \pmeq\ and 
\mmeq,
and $(1,2), (3,2)$ in \mpeq\ and \ppeq. Let us start by considering the  $(2,1)$ 
parts 
of \pmeq\ which give the following constraints,
\eqn\relthree{ \eqalign{
& f_{10} + {y\over 2}  {\partial f_{10} \over \partial y} - 2 f_1 +4 f_9 + r^2 f_4
+ {1\over y}( {\partial f_{3} \over \partial y} + {4\over 5} r^2
{\partial f_{5} \over \partial y}) + {y^2\over 2} r^2 f_{11}  = 0, \cr
& f_{10} + {y\over 2}  {\partial f_{10} \over \partial y} - 2 f_1 - r^2 f_4 
+ r {\partial f_{9} \over \partial r} + f_9
- {1\over y}( {\partial f_{3} \over \partial y} + {4\over 5} r^2
{\partial f_{5} \over \partial y}) \cr & +   {y^2\over 2} f_7  + {y^2\over 2} r^2 f_{11} 
 = 0, \cr
& {2\over 5}{r^2\over y} {\partial f_{5} \over \partial y} - {2\over y}
{\partial f_{3} \over \partial y}  - r {\partial f_{9} \over \partial r} - 3 f_9
+ {y^2\over 2} f_7  = 0.
}}
The $(2,3)$ terms give the equations,
\eqn\relfour{
\eqalign{
& f_{10} + {y\over 2}  {\partial f_{10} \over \partial y} - 2 f_2 - 4 f_8 + r^2 
f_4
+ {1\over y}( {\partial f_{3} \over \partial y} + {4\over 5} r^2
{\partial f_{5} \over \partial y})  + {y^2\over 2} r^2 f_{11}  = 0, \cr
& f_{10} + {y\over 2}  {\partial f_{10} \over \partial y} - 2 f_2 - r
{\partial f_{8} \over \partial r} - f_8 - r^2 f_4
- {1\over y}( {\partial f_{3} \over \partial y} + {4\over 5} r^2
{\partial f_{5} \over \partial y}) \cr & + {y^2\over 2} f_6  + {y^2\over 2} r^2 f_{11} 
 = 0, \cr
& {2\over 5}{ r^2\over y} {\partial f_{5} \over \partial y} - {2\over y}
{\partial f_{3} \over \partial y} + r {\partial f_{8} \over \partial r} + 3 f_8
+ {y^2\over 2} f_6  = 0.
}}
Note that equations \relfour\ are the same as \relthree\ under the 
identification
\ident. This is a nice check that the equations are correct. 

We next need the $(2,1)$ parts of \mmeq\ which give the relations,
\eqn\relfive{\eqalign{
& {y^2\over 2} f_{10} + f_3 + {4\over 5} r^2 f_5 + {1\over y}{\partial f_{4} 
\over \partial y} +{y\over 2}{\partial f_{11} \over \partial y} + f_{11}=0,
\cr
& {y^2\over 2} f_{10} - f_3 - {4\over 5} r^2 f_5 - {1\over y}{\partial f_{4} 
\over \partial y} 
+ {1\over r}{\partial f_{7} \over \partial r} + {y^2\over 2} f_9
+{y\over 2}{\partial f_{11} \over \partial y} + f_{11} =0, \cr
& 2 f_3 - {2\over 5} r^2 f_5 - {1\over r} {\partial f_{7} \over \partial r} 
+ {y^2\over 2} f_9 =0.
}}
From the $(2,3)$ components, we find the equations:
\eqn\relsix{\eqalign{
& {y^2\over 2} f_{10} + f_3 + {4\over 5} r^2 f_5 + {1\over y}{\partial f_{4} 
\over \partial y} +{y\over 2}{\partial f_{11} \over \partial y} + f_{11} =0,
\cr
& {y^2\over 2} f_{10} - f_3 - {4\over 5} r^2 f_5 - {1\over y}{\partial f_{4} \over 
\partial y} 
+ {1\over r}{\partial f_{6} \over \partial r} - {y^2\over 2} f_8 
+{y\over 2}{\partial f_{11} \over \partial y} + f_{11}=0, \cr
& 2 f_3 - {2\over 5} r^2 f_5 - {1\over r} {\partial f_{6} \over \partial r} 
- {y^2\over 2} f_8 =0.
}}
These equations are again consistent with \ident.

We now turn to the $(1,2)$ parts of \mpeq\ which imply that,
\eqn\relseven{\eqalign{
& {y^2\over 2 r} {\partial f_{10} \over \partial r} + {1\over r}{\partial f_{3} 
\over \partial r} + {4\over 5} r {\partial f_{5} \over \partial r} + {28\over 5}
f_5 + {1\over 2} y^2 f_4 + 2 f_9 + {y^4\over 4} f_{11} =0, \cr
& {y^2\over 2 r} {\partial f_{10} \over \partial r} - {1\over r}{\partial f_{3} 
\over \partial r}  - {4\over 5} r {\partial f_{5} \over \partial r} - {28\over 
5}
f_5 + {1\over 2} y^2 f_4 + y^2 f_6 \cr & 
+  y {\partial f_{8} \over \partial y} + 2 f_8 - {y^4\over 4} f_{11}
=0, \cr
& {2\over r}{\partial f_{3} \over \partial r}  - {2\over 5} r 
{\partial f_{5} \over \partial r} - {14\over 5} f_5 - y^2 f_6 
 + y {\partial f_{8} \over \partial y} + 2 f_8 - 2 f_9 =0.
}}
The $(3,2)$ forms give the following set of 
equations:
\eqn\releight{\eqalign{
& {y^2\over 2 r} {\partial f_{10} \over \partial r} + {1\over r}{\partial f_{3} 
\over \partial r} + {4\over 5} r {\partial f_{5} \over \partial r} + {28\over 5}
f_5 + {1\over 2} y^2 f_4  - 2 f_8 + {y^4\over 4} f_{11} =0, \cr
& {y^2\over 2 r} {\partial f_{10} \over \partial r} - {1\over r}{\partial f_{3} 
\over \partial r} - {4\over 5} r {\partial f_{5} \over \partial r} - {28\over 5}
f_5 + {1\over 2} y^2 f_4 + y^2 f_7 -  
y {\partial f_{9} \over \partial y} - \cr &  2 f_9- {y^4\over 4} f_{11}
=0, \cr
& {2\over r}{\partial f_{3} \over \partial r} - {2\over 5} r 
{\partial f_{5} \over \partial r} - {14\over 5}
f_5 - y^2 f_7  -
y {\partial f_{9} \over \partial y} - 2 f_9 + 2 f_8 =0.
}}
Again, \releight\ and \relseven\ are identical under \ident.

The $(1,2)$ parts of \ppeq\ give the following equations, 
\eqn\relnine{\eqalign{
& {y^4\over 4} f_{10} + {y^2\over 2} f_3 + {2 y^2 r^2 \over 5} f_5
+ r {\partial f_{4} \over \partial r}+ 5 f_4 +2f_7 + {y^2\over 2} f_{11} 
+  {y^2 r\over 2}  {\partial f_{11} \over \partial r}
=  0, \cr
& {y^4\over 4} f_{10} - {y^2\over 2} f_3  - { 2 y^2 r^2 \over 5} f_5 
+ r {\partial f_{4} \over \partial r}+ 5 f_4+
y {\partial f_{6} \over \partial y}+ 2f_6 \cr & + r^2 y^2 f_8 - {y^2\over 2} f_{11} 
-  {y^2 r\over 2}  {\partial f_{11} \over \partial r} =  0, \cr
& y^2 f_3  - {y^2 r^2 \over 5} f_5 - y  {\partial f_{6} \over \partial y} 
- 2 f_6 + 2 f_7 + r^2 y^2 f_8 + y^2 f_{11} =0.}}
Lastly, the $(3,2)$ parts give the equations:
\eqn\relten{\eqalign{
& {y^4\over 4} f_{10} + {y^2\over 2} f_3 + {2 y^2 r^2 \over 5} f_5
+ r {\partial f_{4} \over \partial r}+ 5 f_4 + 2f_6+ {y^2\over 2} f_{11} 
+  {y^2 r\over 2}  {\partial f_{11} \over \partial r}
=  0, \cr
& {y^4\over 4} f_{10} - {y^2\over 2} f_3  - {2 y^2 r^2 \over 5} f_5 
+ r {\partial f_{4} \over \partial r} + 5 f_4 +
y {\partial f_{7} \over \partial y}+ 2f_7  \cr & - r^2 y^2 f_9 - {y^2\over 2} f_{11} 
-  {y^2 r\over 2}  {\partial f_{11} \over \partial r}=  0, \cr
& y^2 f_3 - {y^2 r^2 \over 5} f_5 
- y  {\partial f_{7} \over \partial y} + 2 f_6 - 2 f_7  - r^2 y^2 f_9 + y^2 
f_{11} =0.}}
Again, note that \relnine\ and \relten\ are identical under the exchange \ident.

\newsec{The Structure of the Bound State Wavefunction}

\subsec{Reducing the number of functions}

Initially, we have eleven independent functions obeying a set of coupled 
differential equations.
To make progress, we need to reduce the number of functions in a systematic 
fashion.
We can begin to whittle down the number of independent
functions in the following way: the difference of the third equations of 
\relfive\ and \relsix\
together with \relone\ imply that,
\eqn\preconone{f_1=f_2.}
Let us now turn to \relone. By taking 
$\partial_y$ of the second equation and $\partial_r$ of the fourth equation, we 
find
two equations for $ \partial^2_{ry} f_7$. For these equations to be compatible, 
we require
that:
\eqn\conone{ f_{10} - f_1 + f_9 + {1\over y}{\partial f_{11}\over \partial 
y}=0.}
This equation is actually not a new addition to our list of constraints. 
It follows from \relfive\ and \relone. The same analysis applied to \reltwo\ 
gives,
\eqn\contwo{ f_{10} - f_1 - f_8 + {1\over y}{\partial f_{11}\over \partial y} 
=0.}
These equations together require that,
\eqn\contwo{f_8 = - f_9.}
The difference of the fourth equations in \relone\ and \reltwo\ then implies the 
equivalence,
\eqn\conthree{f_6 = f_7.}
In this way, we are reduced to eight functions $ \{ 
f_1,f_3,f_4,f_5,f_7,f_9,f_{10}, f_{11} \}$. 
Half the equations we derived are now redundant since the symmetry \ident\ is an 
actual identity. 

We obtain an algebraic relation between the remaining functions by using the 
third equation
in \relfive\ together with \relone:
\eqn\algebraic{f_3 + {y^2\over 2} f_1 - {r^2\over 5} f_5 + f_{11}=0.}
Using the algebraic relation, we can eliminate one function. The remaining $7$ functions 
of $2$ variables are constrained by $14$ equations, which is the minimal number we could 
have expected. 

\subsec{A first reduction of the equations}

Let us summarize the equations that describe the bound state. We have taken some simpler linear 
combinations of the previous equations:
\eqn\longlist{\eqalign{ (1) & \quad
 r {\partial f_9 \over \partial r}  - 
y {\partial f_1 \over \partial y} - {y^2\over 2} f_7 + 5 f_9 -4 f_1 + 2 f_{10} 
=0, \cr
 (2) & \quad {\partial f_7 \over \partial r} +  r y^2 \left\{f_1 - {f_9\over 2} \right\} + 
2 r f_{11}  = 0, \cr
(3) & \quad r {\partial f_9 \over \partial y} + y {\partial f_1 \over \partial r} + r y f_7 
 =0, \cr
(4) & \quad {\partial f_7 \over \partial y} + r^2 y f_9 + {y^3\over 2} f_1  =0, \cr
(5) & \quad 2 f_{10} + y {\partial f_{10} \over \partial y} - 4 f_1 +5 f_9 
+ y^2 r^2 f_{11} + r {\partial f_{9} \over \partial r} +  {y^2\over 2} f_7= 0, \cr
(6) & \quad 2 r^2 y f_4 -  {y^3\over 2} f_7 + 3 y f_9
- r y {\partial f_{9} \over \partial r}
+ 2 {\partial f_{3} \over \partial y} + {8\over 5} r^2
{\partial f_{5} \over \partial y} =0, \cr 
(7) & \quad {2\over 5}{r^2\over y} {\partial f_{5} \over \partial y} - {2\over y}
{\partial f_{3} \over \partial y}  - r {\partial f_{9} \over \partial r} - 3 f_9
+ {y^2\over 2} f_7  = 0, \cr
(8) & \quad y^2 f_{10} + {y^2\over 2} f_9 + 2  f_{11} + y {\partial f_{11} 
\over \partial y} +  {1\over r}{\partial f_{7} \over \partial r}  =0, \cr
(9) & \quad 2 y f_3 +  {8\over 5} r^2 y f_5 - {y^3\over 2} f_9 + 2 {\partial f_{4} 
\over \partial y} 
- {y\over r}{\partial f_{7} \over \partial r} =0, \cr
(10) & \quad r y (f_4 + f_7) - r {\partial f_{9} \over \partial y} + 
y  {\partial f_{10} \over \partial r} =0, \cr
(11) & \quad {56 \over 5} r f_5 - r y^2 f_7 + 4 r f_9 + {1\over 2} r y^4 f_{11}
 + r y  {\partial f_{9} \over \partial y} + 2 {\partial f_{3} 
\over \partial r} + {8\over 5} r^2 {\partial f_{5} \over \partial r} =0, \cr
(12) & \quad {2\over r}{\partial f_{3} \over \partial r}  - {2\over 5} r 
{\partial f_{5} \over \partial r} - {14\over 5} f_5 - y^2 f_7 
 - y {\partial f_{9} \over \partial y}  - 4 f_9 =0, \cr
(13) & \quad 10 f_4 + 4 f_7 - r^2 y^2 f_9 + {y^4\over 2} f_{10} + 
y {\partial f_{7} \over \partial y} + 2 r {\partial f_{4} \over \partial r}=0, \cr
(14) & \quad y f_3 + {4\over 5} r^2 y f_5 + r^2 y f_9 + y f_{11} - 
 {\partial f_{7} \over \partial y} + r y  {\partial f_{11} \over \partial r}=  0. \cr}}
In these equations, we can remove one function using \algebraic. This is a complicated
set of coupled equations. To uncover the structure of the wavefunction, let us
begin by simplifying as much as possible. 

After staring at these equations for sometime, 
a pair of equations $(3)$ and $(4)$ in \longlist\ appear distinguished. At our special
point, these equations
involve only the $(4,0)$ and $(3,1)$ forms given explicitly in Appendix C. They do not involve
any $(2,2)$ forms. These are the 
analogues of the `top forms' which played a crucial role in proving non-renormalization
theorems \rpss. We can eliminate $f_9$ from these two equations giving, 
\eqn\source{ \eqalign{ \left\{ - \partial_y^2 + {1\over y} \partial_y + r^2 y^2 \right\} f_7
& = y^2 \left\{ 1 + {y\over 2} \partial_y - r \partial_r \right\} f_1, \cr
& = y^2 S(f_1). }}
Note that if we set the source term $S(f_1)=0$, then the homogeneous solutions for $f_7$
are $ e^{\pm r y^2/2}$. The plus sign is not normalizable. However, both solutions suffer
from a more serious problem. The ground state wavefunction must be a smooth function. It must
have a convergent Taylor series about the origin. This implies that each $f_i$ must be function
of $r^2$ and $y^2$ near the origin. The homogeneous solutions alone are therefore ruled out.  

What this teaches us is that $f_7$ is determined in terms of $f_1$. It is convenient to 
make the following redefinition, 
$$ f_7 = y^2 {\tilde f_7}.$$
Note that ${\tilde f_7}$ can have a $1/y^2$ term near the origin. The equation \source\ now
takes the form, 
\eqn\secsource{ \eqalign{ \left\{ - \partial_y^2 - {3\over y} \partial_y + r^2 y^2 \right\} 
\tilde{f_7} & = S(f_1). }}
The left hand side of \secsource\ is the Hamiltonian for a four-dimensional 
harmonic oscillator! Now there is a very pretty collapse. Equation $(4)$ in \longlist\
determines $f_9$ in terms of $f_1$. Likewise $(1)$ and $(2)$ determine $f_{10}$ and
$f_{11}$. Equation $(10)$ determines $f_4$, while $(14)$ determines $f_3$. 
The algebraic constraint \algebraic\ then fixes $f_5$. Clearly, there
many other ways to collapse the problem. The main point is that all the remaining
functions are given in terms of $f_1$. This leaves $8$ equations which must determine
$f_1$. 

\subsec{Solving for $f_7$}

We can now express $f_7$ in terms of $f_1$ in the following way:
\eqn\solvefseven{f_7 = f_7^0(r^2) e^{-ry^2/2} + y^2   
\left\{ - \partial_y^2 - {3\over y} \partial_y + r^2 y^2 \right\}^{-1} S(f_1).}
The $y=0$ component $f_7^0$ is determined by requiring that $f_7$ be smooth, as we
discussed previously. Smoothness of $f_7$ also requires that $f_7^0$ be a function of $r^2$. 
In turn, we can expand $f_1$ as follows:
\eqn\expand{f_1 = \sum_{n=0}^\infty f_1^n(r) \, |n>.}
The $|n>$ are radial harmonic oscillator eigenstates  which obey,
$$  \left\{ - \partial_y^2 - {3\over y} \partial_y + r^2 y^2 \right\} |n> = E_{n} |n>, $$
where $E_{n} = 4(n+1)r$. The construction and properties of these eigenstates are described
in Appendix D. These eigenstates have the nice feature that acting on $|n>$, the operators
$\partial_r, y \partial_y$ and $y^2$ involve only $|n-1>, |n>, |n+1>$. 
Using the relations from Appendix D, we see that the source term has a beautifully simple
form:
\eqn\coefsource{ S(f_1) = \sum_n \left(f_1^n  - r \partial_r f_1^n \right) |n>.}
It is now easy to solve for $f_7$ in terms of $|n>$, 
\eqn\sevensolve{f_7 = f_7^0(r^2) |0> + y^2 \sum_{n=0}^\infty {1\over E_n} 
\left(f_1^n  - r \partial_r f_1^n \right) |n>.}
We have left $y^2$ in \sevensolve\ for later convenience. By considering the coefficient
of $y^2$ in \sevensolve\ and imposing smoothness, we obtain the following relation:
\eqn\sevenrel{ -{r\over 2} f_7^0 + \sum_{n=0}^\infty {1\over E_n} 
\left( 1 - r \partial_r \right) f_1^n =0. }
Since $f_7^0$ is non-singular as $r\r 0$, we obtain the sum rule
\eqn\sumrule{ \sum_{n=0}^\infty {1\over 4(n+1)} f_1^n(0) =0, }
and the relation:
\eqn\invertzero{  f_7^0 = -{2\over r^2} \sum_{n=0}^\infty {1\over 4(n+1)} 
\left( 1 - r \partial_r \right)f_1^n.}
Note that these sum rules may be largely formal since we do not know whether the sums
are absolutely convergent. 

\subsec{Equations for the physics near the flat directions}

Instead of proceeding to reduce the number of functions, let us take a different tack. 
The most interesting physics in this problem occurs in a neighbourhood of the flat 
directions. So let us consider
a Taylor expansion about the flat direction $y=0$. This approach turns out to be more useful 
than reducing the number of functions, which increases
the complexity of the resulting equations. We expand each $f_i$ in the
following way,
\eqn\taylor{ f_i = t_i^0 (r^2) + t_i^2 (r^2) y^2 + t_i^4(r^2) y^4 + \ldots.}
The algebraic constraint \algebraic\ together with the equations of \longlist\ give 
the following set of relations on the $t_i^0$:
\eqn\shortzero{ \eqalign{
(1) & \quad t_3^0 - {r^2\over 5} t_5^0 + t_{11}^0 =0, \cr
(2) & \quad r \partial_r t_9^0 + 5 t_9^0 - 4 t_1^0 + 2 t_{10}^0 =0, \cr
(3) & \quad \partial_r t_7^0 + 2 r t_{11}^0 =0, \cr
(4) & \quad {56\over 5} r t_5^0 + 4 r t_9^0 + 2 \partial_r t_3^0 + {8\over 5} r^2 
\partial_r t_5^0 =0, \cr
(5) & \quad {2\over r} \partial_r t_3^0 -{2\over 5} r \partial_r t_5^0 -{14\over 5} t_5^0
- 4 t_9^0 =0, \cr
(6) & \quad 10 t_4^0 + 4 t_7^0 + 2 r \partial_r t_4^0 =0, \cr
(7) & \quad t_4^0 + 2 t_7^0 + {1\over r} \partial_r \left\{ t_1^0 + t_{10}^0 \right\} =0.
}}
This gives $7$ equations for $8$ unknown functions. As we might have expected, this is not 
sufficient to determine the flat direction physics without input from higher $y$ terms.  
We can similarly derive equations involving only $t_i^2$ and $t_i^0$ which are given in
Appendix E. 

Note the critical observation that we can solve for all $t_i^2$ in terms of the $t_i^0$ using 
$8$ of the equations of $E.1$. There are no new independent functions at order $y^2$ in the
Taylor expansion. We can express the $t_i^2$ in terms of the $t_i^0$ and at most their
second derivatives using the first 
$7$ equations and equation $(10)$ of ${ E.1}$,
\eqn\solvefortwo{ \eqalign{
(1) & \quad   t_7^2 = - {1\over 2} r^2 t_9^0,  \cr
(2) & \quad   t_9^2 = - {1\over 2 r} \left\{ \partial_r t_1^0 + r t_7^0\right\}, \cr
(3) & \quad t_{11}^2 =  {1\over 4}\left\{ 3 t_9^0 + r \partial_r t_9^0 - 2 t_1^0
\right\}, \cr
(4) & \quad t_{5}^2 = - {1\over 2 r}\left\{ r t_4^0 - \partial_r t_9^0 \right\}, \cr
(5) & \quad t_3^2 = {1\over 20}\left\{ -2 r^2 t_4^0 -15 t_9^0 - 3 r\partial_r t_9^0 
\right\}, \cr
(6) & \quad t_4^2 =- {1\over 20 r}\left\{ 10 r t_3^0 + 8 r^3 t_5^0 -5 \partial_r t_7^0   
\right\}, \cr
(7) & \quad t_{10}^2 = {1\over 16} \left\{ -6 r^2 t_{11}^0 + {4\over r} \partial_r t_1^0
+ r \partial_r t_7^0 + \partial_r^2 t_1^0  \right\}, \cr
(8) & \quad t_1^2 =  -{1\over 16 r} \left\{ 8 r t_7^0 + 2 r^3 t_{11}^0 + 4 \partial_r t_1^0
+ r^2 \partial_r t_7^0 + r \partial_r^2 t_1^0  \right\}. \cr }} 
\noindent What we need to check is whether any of the remaining $7$ equations of ${ E.1}$ are
new relations. After some algebra that we will spare the reader, it turns out that all the
remaining equations of ${ E.1}$ are consequences of \shortzero. This is not too surprising. 
The interactions involve order $y^4$ terms so we should not be able to completely determine
the physics on the flat directions without expanding to higher order in $y$. 
So let us expand to order
$y^4$ giving equations relating the $t_i^4, t_i^2$ and $t_i^0$. These equations are again given
explicitly in Appendix E. 

The equations of ${ E.2}$ are very similar to ${ E.1}$, except for mixing with 
some $t_i^0$ terms through
the $y^3$ and $y^4$ interactions in \longlist. It is easy to see that we can again solve for
all the $t_i^4$ in terms of the $t_i^0$. It should be clear that {\it all} the coefficients 
of the higher $y^{2n}$ terms in the Taylor expansion are determined in terms of the $t_i^0$.
The problem is then to determine the $t_i^0$, and we need one more relation in addition to 
those of \shortzero. Again using the first $7$ equations and equation $(10)$ of ${ E.2}$, 
we can solve for the $t_i^4$:
\eqn\solveforfour{ \eqalign{
(1) & \quad   t_7^4 = - {1\over 4} \left\{  r^2 t_9^2 + {1\over 2} t_1^0 \right\}, \cr
(2) & \quad   t_9^4 = - {1\over 4 r} \left\{ \partial_r t_1^2 + r t_7^2\right\}, \cr
(3) & \quad t_{11}^4 =  {1\over 16}\left\{ 8 t_9^2 +2 r \partial_r t_9^2 - 8 t_1^2
+ {1\over r} \partial_r t_1^0 \right\}, \cr
(4) & \quad t_{5}^4 =  {1\over 16 r^3}\left\{ r t_7^0 - 4 r^3 t_4^2 + 2 r t_9^2
+ \partial_r t_1^0 + 4 r^2 \partial_r t_9^2 \right\}, \cr
(5) & \quad t_3^4 = {1\over 80 r}\left\{  r t_7^0 - 4 r^3 t_4^2 - 38 r t_9^2
- 4 \partial_r t_1^0 -6 r^2 \partial_r t_9^2 \right\}, \cr
(6) & \quad t_4^4 = {1\over 80}\left\{ 5 t_9^0 - 20 t_3^2 -16 r^2 t_5^2 
+{10\over r} \partial_r t_7^2 \right\}, \cr
(7) & \quad t_{10}^4 = {1\over 200} \left\{ 20 r^2 t_1^0 + 40 r^2 t_3^2 -8 r^4 t_5^2 -5 t_7^2
+ {20\over r} \partial_r t_1^2 + 5 r \partial_r t_7^2 + 5 \partial_r^2 t_1^2 \right\}, \cr
(8) & \quad t_1^4 =  -{1\over 200 r} \big\{ -5 r^3 t_1^0 - 10 r^3 t_3^2 + 2 r^5 t_5^2
+ 45 r t_7^2 + 20 \partial_r t_1^2 \cr & \quad\qquad 
+ 5 r^2 \partial_r t_7^2 + 5 r \partial_r^2 t_1^2
\big\}. \cr }} 
Note that the expression for $t_7^4$ agrees with the 
expression coming from the earlier relation \source\ that we derived between $f_1$ and $f_7$. Once again we are left with $7$ additional equations. It turns out that these additional 
equations again give no new relations. After checking higher order Taylor coefficients, 
we find that
there are no further relations. It appears that any choice of $t_i^0$ satisfying 
\shortzero\ give a zero energy
solution. However, most of these solutions are not normalizable. 

It is straightforward to derive
the general recursion relation for $t_i^n$ in terms of lower Taylor coefficients:
\eqn\generaltay{ \eqalign{
(1) & \quad   t_7^{n} = - {1\over n} \left\{  r^2 t_9^{n-2} + 
{1\over 2} t_1^{n-4} \right\}, \cr
(2) & \quad   t_9^n = - {1\over  n r} \left\{ \partial_r t_1^{n-2} + r t_7^{n-2}\right\}, \cr
(3) & \quad t_{11}^n =  {1\over 2}  \left\{  {1\over 2} t_9^{n-2} - t_1^{n-2} - {1\over r} 
\partial_r t_7^n \right\}, \cr
(4) & \quad t_{5}^n =  {1\over n}\left\{ {1\over r} \partial_r t_9^{n-2} - t_4^{n-2} 
\right\}, \cr
(5) & \quad t_3^n = \left\{  {r^2 \over 5} t_5^n - t_{11}^n - {1\over 2} t_1^{n-2}\right\}, \cr
(6) & \quad t_4^n = {1\over 2 n}\left\{ {1\over r} \partial_r t_7^{n-2} + {1\over 2} t_9^{n-4}
- 2 t_3^{n-2} -{8 \over 5} r^2 t_5^{n-2} \right\}, \cr
(7) & \quad t_{10}^n = - {1\over 2 n (n+6)} \left\{ (n +8 ) t_7^{n-2} + {10 n} 
t_9^n + r^2 ( 2 n +8) t_{11}^{n-2} + {2 n} r \partial_r t_9^n \right\}, \cr
(8) & \quad t_1^n = -{1\over n } \big\{ t_7^{n-2} + n  t_{10}^n + r^2 t_{11}^{n-2} 
 \big\}. \cr }} 
Let us close this discussion of the Taylor expansion by pointing out that the entire 
Taylor series depends only on $3$ of the $t_i^0$. 
To see this, let us express \shortzero\ in a form which will be more convenient for 
later manipulation,
\eqn\newzero{ \eqalign{
(1) & \quad t_5^0 = {5\over r^2} \left( t_3^0 + t_{11}^0 \right), \cr
(2) & \quad t_{11}^0 = - {1\over 2 r} \partial_r t_7^0, \cr
(3) & \quad t_{7}^0 = - {1\over 2} \left( 5 t_4^0 + r \partial_r t_4^0 \right), \cr
(4) & \quad  t_9^0 = {5\over 6 r} \partial_r t_3^0, \cr
(5) & \quad t_{10}^0 = 2 t_1^0 - {5\over 2} t_9^0 - {r\over 2} \partial_r t_9^0, \cr
(6) & \quad  20 \, \partial_r ( r^3 t_3^0)  + r \left\{ 72 \partial_r  
+ 33 r \partial_r^2  + 3 r^2 \partial_r^3 \right\} t_4^0 =0, \cr
(7) & \quad \partial_r t_1^0 + {1\over 48 r^4} \{ -64 r^5 + 72 \partial_r 
-16 r^6 \partial_r - 72 r \partial_r^2 \cr & \quad +12 r^2 \partial_r^3 + 12 r^3 \partial_r^4 
+ r^4 \partial_r^5 \, \} \, t_4^0 =0. \cr }}
The first $5$ equations of \newzero\ are definitions of various $t_i^0$. 
The last $2$ are relations on the $3$ independent functions $t_1^0, t_3^0$ and $t_4^0$. 

\newsec{An Asymptotic Expansion of the Wavefunction}
\subsec{Matching Taylor and oscillator expansions}

It is hard to see how to implement the normalizability condition in a Taylor series. So although
we have found the zero energy solution in terms of the $t_i^0$, we need a practical procedure
to construct the $t_i^0$. We will determine the $t_i^0$ under the assumption that the 
bound state wavefunction admit an asymptotic expansion in powers of $1/r$. 
The primary motivation for studying the asymptotic
expansion is that the asymptotic form should be interpretable in terms of supergravity 
plus higher derivative corrections. What we do not know is whether the
asymptotic expansion converges to the actual bound state wavefunction. 
This issue is
closely related to the following two questions: if we sum the effects of all higher
derivative terms in the effective action for this gauge theory, does the result give 
non-singular physics
as $r \r 0$? If we sum the effects of all higher derivative terms beyond supergravity 
on the spacetime solution for a $5$-brane in M theory, do we find a smooth convergent 
solution near the $5$-brane?

These questions are intrinsically tied to the problem of gauge-fixing the Taylor series 
solution in a way that results in a normalizable solution. We want to construct the $t_i^0$ 
in a useful systematic expansion. The dominant terms in an asymptotic expansion 
are those that decay polynomially in $1/r$.  We point out that for large $r$, 
an approximate asymptotic bound state can be constructed in a $1/r$ expansion using a 
method described in \rsavmark. An analytic expansion in $1/r$ near infinity is essentially
a perturbative expansion, although not in $g^2$ a priori but in $g^{2/3}$. 
A similar technique was used in \refs{ \rhs, \rpw, 
\rfrolich} to further explore the long distance dynamics and the
asymptotic structure of the bound state wavefunction for $2$ D0-branes. Unfortunately, the
effective long distance Hamiltonian has only been constructed to order $1/r^2$, 
which is the required order for an index computation \refs{\rsavmark, \ryi}. With our
knowledge of the Taylor series solution \generaltay\ for this problem, we can do significantly
better than those approximate constructions. 

Can there be non-perturbative terms? These are terms which are not visible
in a $1/r$ expansion, like $e^{-r^2}$, but which become important as $r$ 
becomes small.   We do not actually know whether there
are any such terms, and there are no candidates like instanton configurations that could
generate these terms in the abelian gauge theory. This leads the first author 
to suspect that the 
analytic expansion in $1/r$ might well be exact. Nevertheless, we cannot prove that 
non-perturbative terms are not present. To really rule out such terms, we need a technique 
for finding the 
bound state solution which is inherently more global than an asymptotic expansion.
We shall discuss a more global approach in section $6$.

What is hard to see in the Taylor expansion of section $4$
is normalizability in the $y$-direction. This is much easier to see in the oscillator
expansion so we will match the two expansions to get control over the question of 
normalizability. When
expanded in a harmonic oscillator basis with frequency proportional to $r$, 
\eqn\nextexp{ f_i = \sum_{n=0}^\infty f_i^n(r) \, |n>,}
the $f_i^n$ must decay as $r \r \infty$.
Let us expand each $ f_i^n$ in powers of $1/r$. This 
implies that the $t_i^n$ have an expansion in $1/r$. We can then reorganize the Taylor 
series for $f_i$ in the following way: 
\eqn\organize{ \eqalign{ f_i & = t_i^0 + t_i^2 y^2 + t_i^4 y^4 + \ldots, \cr
& =  \sum_p {1 \over r^p} \, \sum_k \, b_k^{p, i} \, |k>.}}
The $  b_k^{p, i}$ are just some numbers which determine the
collection of harmonics contributing to a given power $1/r^p$.  
Note that the oscillator eigenstates depend 
only on the combination $r y^2$ so that $ \sum_k b_k^{p, i} |k> $ is a power series
in $ry^2$. 


For example, suppose that
the harmonic $|m>$ is the only harmonic with a non-zero coefficient in the 
sum  $ \sum_k b_k^{p, i} |k> $ for the  $1/r^p$ term of $f_i$. 
We stress that in general there
can be many harmonics contributing to a given term, but for simplicity, let us assume there
is just one. The Taylor series for $f_i$ must then contain the terms:
\eqn\dominant{ \eqalign{ f_i = & \, b_m^{p, i} \,  {1\over r^p} 
\left( 1+a_1^{(m)} y^2 + \ldots 
+ a_m^{(m)} y^{2m} \right) e^{-ry^2/2} + \ldots, \cr
= &  \, b_m^{p, i} \, {1\over r^p}  \left( 1 +  \left\{ a_1^{(m)} - 
{1\over 2} r \right\} y^2 + \left\{ 
a_2^{(m)} - {1\over 2} r a_1^{(m)} + {1\over 2^2 2!} r^2 \right\} y^4 + \ldots \right)
+ \ldots,\cr
= & \, b_m^{p, i} \, {1\over r^p} \left( 1 - {1\over 2}(1+m) r y^2 + 
{1\over 24}(3+4 m+2 m^2 ) r^2 y^4 + \ldots \right) + \ldots. } }
There are specific relations between the Taylor coefficients in 
\dominant.  We want to impose relations of this kind on the $t_i$ to satisfy  
normalizability in the $y$-direction for each choice of $p$ in the $1/r$ expansion. 

\subsec{The structure of the solution}

Our Taylor series solution is completely determined by the three functions $t_1^0, t_3^0$
and $t_4^0$. These three functions must obey equations $(6)$ and $(7)$ of \newzero. 
To ensure that $t_4^0$ is decaying, we see from $(6)$ that $t_3^0$ must
take the form,
\eqn\leadtthree{ t_3^0 = {c_1 \over r^3} + \ldots, }
where omitted terms decay more rapidly. For a given choice of $p$ in the $1/r$ expansion, 
we wish to extract the terms in  each $t_3^n$ which 
contribute to $ \sum_k b_k^{p, 3} |k> $. Again, this is just the 
statement that we can organize the Taylor series for $f_3$ so that,
\eqn\partform{ f_3 = {1 \over r^3} \sum_k b_k^{3, 3} |k> + \ldots, }
for some coefficients $b_k^{3, 3}$. This is formally true for any Taylor series. 
What is generally not
true is that generic $b_k^{3, 3}$ give a wavefunction normalizable in the $y$
direction. Heuristically, the norm of $f_3$ should be dominated by the leading term in the
$1/r$ expansion. If we compute the norm of $f_3$ under this assumption, we see that:
$$ |f_3|^2 \sim \int \, r^4 dr \left\{ \left( {c_1 \over r^3} \right)^2 {1\over r^2} 
\sum_k {|b_k^{3, 3}|^2 \over (2+2k)} \right\}, $$
where we have taken the normalization of $|k>$ given in Appendix D. 

By looking at \solvefortwo, \solveforfour, \generaltay\ and \newzero, we can see what terms
in the Taylor series for $f_3$ have the right structure to determine the $b_k^{3, 3}$ 
coefficients
of \partform. To quickly answer this question, let us list some order of magnitude 
relations which 
follow from \newzero\ for the perturbative expansion. All relations
are given in terms of our $3$ independent functions $t_1^0, t_3^0$ and $t_4^0$, 
\eqn\orderone{ \eqalign{
(1) & \quad t_5^0 \sim O( t_3^0/r^2 + t_4^0/r^4), \cr
(2) & \quad t_{11}^0 \sim O(t_4^0/r^2), \cr
(3) & \quad t_{7}^0 \sim O(t_4^0), \cr
(4) & \quad  t_9^0 \sim O(t_3^0/r^2), \cr
(5) & \quad t_{10}^0 \sim O(t_1^0 + t_3^0/r^2). \cr }}
There are similar order of magnitude relations for $t_i^2$ from \solvefortwo,
\eqn\ordertwo{ \eqalign{
(1) & \quad   t_7^2 \sim O(t_3^0), \cr
(2) & \quad   t_9^2 \sim O(t_1^0/r^2 + t_4^0), \cr
(3) & \quad t_{11}^2 \sim O(t_3^0/r^2 + t_1^0), \cr
(4) & \quad t_{5}^2 \sim O(t_4^0 + t_3^0/r^4), \cr
(5) & \quad t_3^2 \sim O(r^2 t_4^0 + t_3^0/r^2), \cr
(6) & \quad t_4^2 \sim O(t_3^0 + t_4^0/r^2), \cr
(7) & \quad t_{10}^2 \sim O(t_4^0 + t_1^0/r^2), \cr
(8) & \quad t_1^2 \sim O(t_4^0 + t_1^0/r^2). \cr}}
It is easy to continue and list order of magnitude relations for $t_i^4$ 
from \solveforfour, and higher $t_i^n$ using \generaltay. These relations are 
useful for easily determining which terms in $t_i^n$ are relevant for determining the 
oscillator coefficients $b_k^{p, i}$. 

Returning to our specific case of $f_3$, we see that since $t_3^0 \sim 1/r^3$, 
the only terms in $t_3^2$ relevant for computing $b_k^{3, 3}$ 
are those proportional to $ 1/r^2$.  
Looking at $t_3^2$ from \ordertwo, we see that
the only way to have a term contributing to the $ \sum b_k^{3, 3} |k>$ 
is if, 
\eqn\couldbe{t_4^0 = {c_1' \over r^4} + \ldots.}
It could be the case
that $c_1'=0$, which means that the $b_k^{3, 3}$ sum up in such a way that the $y^2/r^2$
term in \partform\ vanishes. Actually if $c_1'=0$, the situation is much worse: we can see 
after
some work that the $b_k^{3, 3}$ have to be chosen so that the $ (ry^2)^{1+2n}$ terms 
in the Taylor expansion of, 
$$ \sum_k b_k^{3, 3} |k>, $$
vanish for all $n$. The  $ (ry^2)^{2n}$ terms are all proportional to $c_1$ and give the 
constraint, 
\eqn\nicecoef{ c_1 \left( 1 + {1\over 8} (ry^2)^2 + {1\over 348}  (ry^2)^4 + {1\over 46080} (ry^2)^6
 + \ldots \right) =  \sum_k b_k^{3, 3} |k>. }
However, the left hand side of \nicecoef\ is the expansion of the 
lowest oscillator $|0>$ with the  $ (ry^2)^{1+2n}$ terms missing. Since $|0> = e^{-r y^2/2}$,
we immediately see that the left hand side of \nicecoef\ sums to the expression:
$$ {1\over 2} \left\{  e^{-r y^2/2} + e^{ r y^2/2}\right\}. $$ 
This is clearly not normalizable so $c_1'$ is necessarily non-zero. We will use this kind
of argument repeatedly to determine the asymptotic solution. 

With $c_1' \neq 0$, we get a modified constraint:
\eqn\bettercoef{ c_1 \left( 1 + {1\over 8} (ry^2)^2
 + \ldots \right) - c_1' \left( {1\over 10}(ry^2) + {1\over 240} (ry^2)^3 + \ldots  \right) 
=  \sum_k b_k^{3, 3} |k>. }
It is easy to see that the terms proportional to $c_1'$ sum up to give,\foot{To see
this, let us introduce new variables $(r, u= ry^2/2)$. In terms of 
these variables and 
simple redefinitions of the $f_i$, the equations of \longlist\ can be put in a
triangular form with respect to the grading induced by the $1/r$ expansion. In doing
this, we treat $u$ as independent of $r$. The resulting triangular system is exactly
solvable in the $1/r$ expansion.} 
$$ {1\over 10} \left\{  e^{-r y^2/2} - e^{ r y^2/2}\right\}. $$ 
In this case, it is completely clear that we only get a normalizable solution if
we pick,
\eqn\crelone{ c_1' = 5 c_1. }
More generally, we will encounter the situation where we have a Taylor series of the 
form,
\eqn\sampletay{ \a \sum_n  d_n \, (ry^2)^{2n} + \a' \sum_n  d_n' \, (ry^2)^{2n+1},}
where we know both $d_n$ and $d_n'$, and either $\a$ or $\a'$. 
Let us assume we know $\a$ and 
we wish to determine the values of $\a'$ for which the Taylor series is normalizable. 
If the Taylor series can be fitted by a finite
number of oscillators for some $\a'$ then that choice of $\a'$ is unique. 
To see this, suppose there were 
two distinct choices of $\a'$. We could take the difference between the two series to obtain
a Taylor series proportional to, 
$$ \sum_n  d_n' (ry^2)^{2n+1}. $$
However, this series cannot be generated by any sum over a finite number of
oscillators since all
the $ (ry^2)^{2n}$ terms must vanish. Therefore the choice of $\a'$ is unique.

From equation $(6)$ of \newzero, we see that the $O(1/r^4)$ term in $t_4^0$ 
induces an $O(1/r^6)$ correction to $t_3^0$. This begins to suggest a perturbation
expansion in $1/r^3$, and indeed that seems to be the case. From $t_4^2 \sim O(1/r^6)$, 
we see that we need a $O(1/r^7)$ correction to $t_4^0$ and so on.  It is easy to add
$t_1^0$ into the story: from the expression for $t_1^2 
\sim O(1/r^4)$, we learn that $t_1^0 \sim 1/r^5$ at leading order. From 
$(7)$ of \newzero, we see that $t_1^0$ mixes with the first correction to $t_4^0$
and so on. What we have uncovered is the minimal required form for the solution given
$c_1 \neq 0$. The choice of $c_1$ then determines the normalization of the wavefunction. 
The solution is given by the expansion:
\eqn\solstruct{ \eqalign{
t_3^0 & =  {c_1\over r^3} + {c_2\over r^6} +{c_3\over r^{9}} + \ldots, \cr
 t_4^0 & = {c_1'\over  r^4} + {c_2'\over r^7} + { c_3' \over  r^{10}} + \ldots, \cr
t_1^0 & = {c_1''\over r^5}  + {c_2'' \over r^8 } + { c_3'' \over r^{11}} +\ldots.}}
It is interesting that the $t_i^0$ have an expansion
in powers of $1/r^3$.
Restoring the coupling constant, we see that this is an expansion in $g^2/r^3$. This is the
natural expansion parameter in the  gauge theory.  From the 
perspective of the effective action on the Coulomb branch, what
we are doing is summing the effects of the metric and all higher derivative corrections on
the vacuum state of the gauge theory. 

The expansion \solstruct\ represents the minimal required terms in the solution. It is 
natural to ask whether other powers of $1/r$ are possible. Equation $(6)$ of \newzero\ 
requires that, 
$$ t_4^0 \sim O(r^2 t_3^0), $$
except for the special case where $t_3^0 \sim O(1/r^3)$. Likewise, equation $(7)$ of
\newzero\ requires that, 
$$ t_1^0 \sim O( r^2 t_4^0 + {1\over r^4} t_4^0),$$
again except for the case $t_4^0 \sim O(1/r^4)$. Let us be concrete: suppose 
$t_3^0$ has a term of order $O(1/r^4)$. This requires $t_4^0 \sim O(1/r^2)$ and
therefore $t_1^0 \sim O(1)$, which is not possible. Suppose $t_3^0$ has a term of 
order $O(1/r^5)$, which implies that $t_4^0 \sim O(1/r^3)$. In turn, we see that
$t_3^2 \sim O(1/r)$ which requires that $t_3^0  = \a/r^2 $. This is impossible unless
$\a=0$. In that case, the Taylor expansion of $f_3$ has terms of the schematic form:
$$ f_3 \sim O( {y^2\over r}) + O( r y^6 ) + \ldots.$$
As in our earlier discussion, it is not hard to check that this sum is not normalizable. 
This kind of argument extends to higher powers: suppose $t_3^0$ has a term of 
order $O(1/r^7)$, which implies that $t_4^0 \sim O(1/r^5)$. From $t_3^2 \sim O(1/r^3)$, 
we require a $O(1/r^4)$ correction to $t_3^0$ which brings us back to an earlier case. 
This kind of reasoning suggests that only those terms generated by the original 
$1/r^3$ term in $t_3^0$ appear in the solution. This also 
agrees with our gauge theory intuition.
We shall therefore restrict our attention to the terms of \solstruct. 

\subsec{Supergravity and the leading terms of the solution}
 
So far, we have found that the leading order term of $f_3$ is proportional 
purely to the oscillator ground state $|0>$. We also know all the leading powers for
the $t_i^0$. A glance at Appendix C tells us that 
the dominant terms  for large $r$ in the bound state $\Psi$ are 
$f_3, f_4$ and $f_5$. Note that on the flat directions where $y=0$, the only non-vanishing
forms in $\Psi$ are those with coefficients $f_3, f_4$ and $f_5$.  
It is not hard to check that both
$f_3$ and $f_5$ are also proportional to $|0>$ at leading order.  
Recalling that $r = |x^1|$, we have learnt that these leading
terms are given by, 
\eqn\compsg{ \eqalign{ & f_3   = {c_1 \over r^3} |0> +\ldots, \cr
& x^1 f_4 = {5 c_1 x^1\over r^4} |0> + \ldots, \cr
& (x^1)^2 f_5  = {5 c_1 (x^1)^2 \over r^5} |0> + \ldots. } }
Although each term transforms very differently under $Spin(5)$ in either 
the $\1, \5$ or ${\bf 14}$, each decays at the
same rate at leading order.

We can compare what we have learned about the exact 
Taylor series solution with
what we might expect from an effective Hamiltonian construction like the one given in
\rsavmark. 
To construct the approximate bound state, we note that the wavefunction is sharply localized 
near the flat directions for large $r$. The potential
term \V\ is dominated by the $r^2 y^2$ term. We conclude that for large $r$, the wavefunction
can be expanded in a harmonic oscillator basis, with higher oscillator modes suppressed by 
powers of $1/r$. The approximate bound state wavefunction takes the form of a product:
there is a wavefunction of the light vector multiplet degrees of freedom multiplied by the
ground state for the massive hypermultiplet degrees of freedom. We can now construct the 
effective Hamiltonian governing the long distance physics in a $1/r$ expansion. 
The hypermultiplet ground state 
is easily determined from \hyperalg\ to have the form,\foot{The natural decomposition 
of $x^\mu$
and $q_i$ into light and heavy variables at large $r$ makes the construction of the approximate
ground state much simpler than the non-abelian D0-D0 theory considered in \rsavmark.}
\eqn\approxgnd{ r e^{-r y^2/2} dv_1 dv_2.}
In the most primitive approximation where the effective interactions are cancelled to 
order $1/r$, the effective Hamiltonian is $ (1/2) p^\mu p^\mu $ and acts on 
\approxgnd\ multiplied by some function of $x$ and $du$. In fact, to determine the leading
decay, we should go to order $1/r^2$. Even then there will be a degenerate set of approximate
ground states.  

However, this rough approximation is good enough for the purpose of comparison with the
structure following from the Taylor series solution. Using the forms 
given in Appendix C, we see that the coefficients in \compsg\ precisely
conspire to give agreement with the approximate construction. The leading terms in 
$\Psi$ sum to give,
\eqn\leadingexact{ \eqalign{ \Psi & =  f_3 |2,2>_{\1} +  f_4 x^1 |2,2>^1_{\5} + 
f_5 (x^1)^2  |2,2>^{1 1}_{\bf 14} + \ldots, \cr
& \sim {10 c_1 \over r^3} e^{-r y^2/2} (du_1 du_2-du_3 du_4) (dv_1 dv_2) + \ldots, \cr}}
where we take $x^1$ positive. There are a number of strange features of this asymptotic
solution. First note that the vacuum for the massive fermions $dv_1 dv_2$ is not a single
representation of $Spin(5)$. Let us contrast this with the case of $2$ D0-branes where the 
vacuum for the massive fermions transforms in a single representation, the ${\bf 44}$, 
of $Spin(9)$ \refs{\rhs, \rfrolich}. The appearance of three different representations, the 
$\1, \5$ and $ {\bf 14}$, 
in the asymptotic solution really cries out
for an interpretation both in terms of the DLCQ M theory $5$-brane, and in terms of the 
supergravity solution for the D0-D4 bound state. 

What this seems to suggest is that the massive degrees of freedom never really decouple at 
large $r$. Through the $Spin(5)$ flavor symmetry, the vacuum always knows about the 
existence of massive degrees of freedom. This issue is intimately tied to uniqueness 
of the bound state: the statement that the bound state is unique involves knowledge
about both long and short distance physics. Uniqueness however forces invariance under
the full $Spin(5)$ acting on both light and heavy degrees of freedom. That $dv_1 dv_2$ is 
not an irreducible representation then requires particular combinations of
$Spin(5)$ representations for the light degrees of freedom. A deeper understanding of this 
issue is certainly in order. 

\subsec{Beyond supergravity}

Let us return to our general solution \solstruct. We determined $c_1'$ in terms of 
$c_1$ in \crelone. How do we determine $c_1'' \, $? The straightforward way to determine 
$c_1''$ is not particularly elegant. We can take the 
Taylor series for $f_1$ and impose normalizability in the $y$-direction,
\eqn\ssurprise{ \eqalign{ f_1 & =   \left( {c_1''\over  r^5} \right)  +  
\left( {5 c_1\over 4 r^4} 
\right) y^2 
+  \left( {c_1'' - 10 c_1 \over 40 r^3} \right) y^4 +  
\left( {5 c_1 \over 96 r^2} \right) y^6 \cr
& +  \left( {c_1'' \over 4480 r} - {c_1 \over 168 r} \right) y^8  + O(y^{10}) + \ldots. \cr}}
Note the general structure: half the terms in the Taylor series are known. The other
half depend on the unknown constant that we wish to determine. It is not hard to see
that the $ (ry^2)^{1+2n}$ terms come from expanding the lowest oscillator $|0>$
state. The situation is then essentially the same as in \bettercoef, and we find 
that:
\eqn\creltwo{ c_1'' = - {5\over 2} c_1.}
From equation $(6)$ of \newzero, we find that $c_2$ is determined by $c_1'$:
\eqn\crelthree{ c_2 = c_1. }
Equation $(7)$ of \newzero\ relates $c_2'$ to $c_1''$, 
\eqn\crelfour{ c_2'  = - {25\over 2} c_1.}
We are again back to the question of studying $f_1$ to determine $c_2''$. 
The relevant terms in $f_1$ begin at order $1/r^8$, 
\eqn\foneexpand{ \eqalign{ f_1 =& {1 \over r^8} \bigg( \, c_2'' + {125 c_1 \over 16} (ry^2) 
+ {c_2'' - 25 c_1 \over 40}
(ry^2)^2 + {25 c_1\over 384} (ry^2)^3 \cr & + \left\{ {c_2'' \over 4480} + {5 c_1 \over 384} 
\right\} (ry^2)^4 +O(r y^2)^5  \,
\bigg) + \ldots.} } 
Apart from more complicated coefficients, the general pattern is the same. Half the 
terms in the Taylor series are proportional to $c_1$ while the other half depend on the
unknown constant $c_2''$. We can try to match \foneexpand\ with an oscillator expansion
involving a finite number of oscillators. Indeed, for the choice
\eqn\crelfive{ c_2''  =  - {175 \over 8} c_1, }
we can fit \foneexpand\ by the first two oscillators:
$$ f_1 = - {225 \over 8 }  {c_1\over r^8} \left( |0> - {2\over 9} |1> \right) + \ldots. $$
As we might expect, higher excited oscillator states appear in the solution as we study
more rapidly decreasing terms. This also suggests that only a finite number of oscillators
will appear at any given order in the $1/r$ expansion. 

Let us iterate the argument one more time. From $(6)$ and $(7)$ of \newzero, we immediately 
obtain the following relations:
\eqn\crelsix{ \eqalign{  & c_3 = {35\over 2} c_1, \cr 
& c_3'  = - {195\over 2} c_1. }}
The relevant terms in $f_1$ take the form, 
\eqn\fonetwo{ \eqalign{ f_1 =& {1 \over r^{11}} \bigg( \, c_3'' + {2825 c_1 \over 16} (ry^2) 
+ \left\{ {c_3'' \over 40}  - {595 c_1\over 32} \right\}
(ry^2)^2 \cr & + {425 c_1\over 128} (ry^2)^3 + O(r y^2)^4 \bigg) + \ldots, }}
and can be fit by the first three oscillator modes,
$$ f_1 =  - {525\over 32} {c_1\over r^{11}} 
\left( |2> - {208\over 21} |1> +{115\over 3}|0> \right) + \ldots, $$
for the choice, 
\eqn\crelseven{ c_3''  = - {7725\over 16} c_1. }

\newsec{A Remarkable Reduction}
\subsec{Prolongation}

Studying the bound state solution in an asymptotic series is analogous to 
studying the M5-brane in a derivative expansion around the supergravity
solution. Ideally, we want a more powerful technique to solve the vacuum equations. 
The aim of this final section is to present a more global approach to solving the
vacuum equations. Hopefully, this approach is 
closer to the method an M theorist might use to study the M5-brane.  

We shall present a surprising reduction of the long list of equations \longlist\ 
to a single scalar elliptic equation. The equation takes the form,
\eqn\surtwo{ \left( \Delta + \vec{B} \cdot \nabla + W \right) u =0, }
where $ \Delta = \partial_r^2 + \partial_y^2$, the vector field $ \vec{B}$ 
with components $(B_r, B_y)$ and the potential $W$ are rational functions of $r$ and $y$.
The function $u = u(r,y)$ is a particular combination of the $f_i$. 
 In order to explain how we reduce our first order system to a single second
order equation, let us first consider the
inverse process: prolongation. 
To understand the procedure, let us begin with an illustrative
 example: we take an equation of the 
form,
\eqn\sample{ F_{xx} + F_{yy} + U F =0. }
We can define new functions $p=F_x$ and $q=F_y$ with which we can 
`prolong' our scalar second order equation into a system of first order equations:
\eqn\samone{ \eqalign{  & p_x+q_y+UF =0, \cr
&  F_x  =p, \cr
&  F_y  =q. \cr} }
We also have a compatibility relation,
$$F_{xy}=F_{yx}, $$
which implies an additional fourth equation:
\eqn\sampadd{ p_y-q_x = 0.}
We can express these relations as a differential system,
\eqn\samdiff{ \eqalign{ & dF=pdx+qdy, \cr 
& dp=adx+bdy, \cr
& dq=b dx - (a+U F) dy.} }
We could repeat the procedure and prolong again by adding the equations, 
\eqn\repeat{ \eqalign{ & da=cdx+edy, \cr
& db=edx+((UF)_x-c)dy. \cr}}
Each time we prolong a system of equations like this, we add two new 
unknown functions. These functions are the unknown derivatives of the functions
comprising the previous system of equations.

\subsec{Deprolongation}

Now we would like to `deprolong' our system of equations \longlist.\foot{
We are especially grateful to Robert Bryant for suggesting and explaining this 
reduction to us.} We shall see
that the system \longlist\ containing $7$ independent functions can be obtained
by prolonging a single second order equation three times. In order to deprolong
\longlist, we write  the equations as an exterior system 
as before.  We choose $5$ functions  $F_i$ which are linear combinations of
the initial seven $f_i$ so that $dF_i$ is expressible algebraically in terms of 
the original seven.
We keep the $5$ equations defining the $dF_i$ and discard the remaining two equations.
For example, for this first step,
we can make the following choice:
$$ \eqalign{ F_1  & = f_1 + f_{10}, \cr
F_2  & = r^2 f_5 - y^2 f_1, \cr} $$
$$ F_3 = f_7, \quad F_4  = f_4,  \quad F_5  = f_{11}. $$
So far the equations remain first order.
We then iterate this procedure until we arrive at the first prolongation of the 
scalar equation. We then make the obvious  substitution to 
transform the first prolongation into a second order scalar equation.      

Let us summarize the results of the deprolongation. We obtain the following
equation:
\eqn\prolong{ u_{rr}+u_{yy}+ { \widetilde{B}_r \over r F} \, u_r + { \widetilde{B}_y 
\over y F} \, u_y+ { \widetilde{W}
\over F} \, u=0.}
We define $u$ in terms of the variables $s=y^2/2$ and $t=r^2/2$ and the functions 
$f_1, f_4, f_5, f_7, f_{10}, f_{11}$:
\eqn\defonepro{ \eqalign{
u = & \left( 2s-t \right) f_7 - \left({4\over 3} t s^2-{1\over 3} t^2 s-{1\over 6} t^3-
1 \right) f_{10} 
+  \left( {4\over 3} t s^2 - {1\over 3} t^2 s- {1\over 6} t^3+ 1 \right) f_1 
\cr & - s f_4 + \left( {2\over 3} t s^2-{8\over 3} s^3+ {1\over 3} t^2 s \right)f_5 
- \left( {2\over 3} t s- {8\over 3} s^2+ {5\over 6} t^2 \right)f_{11}.}} 
It would be interesting to find a geometric interpretation for \prolong\ -- perhaps in
terms of some bundle over either $\IR^2$ or $\IR^9$. Let us list 
the rational functions which appear in \prolong. 
For $\widetilde{B}_r$, we find the expression:
\eqn\twopro{ \eqalign{
\widetilde{B}_r= &
-224 t^3 s^3 + 576 t^2 s^4 + 32 t^4 s^2 - 12 t^5 s + 4 t^6 - 12 t^2 s \cr &
+ 144 t s^2 - 24 t^3
- 448 s^5 t-48s^3 +36, \cr}}
while $\widetilde{B}_y$ and $F$ are given by,
\eqn\threepro{
\eqalign{
\widetilde{B}_y = & 272 t^2 s^4 - 192 t^3 s^3
- 40 t^4 s^2 + 264 t s^2 - 216 t^2 s + 68 t^5 s + 42 t^3
- 15 t^6 +9 - 64 s^5 t, \cr
F = & 9 + 64 t^3 s^3 - 8 t^4 s^2 - 6 t^3 + 24 t^2 s - 112 t^2 s^4 + t^6 - 24 t s^2 
+ 64 s^5 t - 4 t^5 s.\cr}}
Lastly, the function $\widetilde{W}$ which determines the potential $W$ takes the form:
\eqn\fourpro{
\eqalign{
\widetilde{W} = & - 105 t^2 + 80 s^3 t^2 + 38 t^5 + 144 s t - 96 t^3 s^2 + 24 s^2  
- 144 t^4 s^4 
- 32 t^5 s^3 \cr & + 24 t^6 s^2 - 256 s^6 t^2 + 224 s^4 t + 384 s^5 t^3 - t^8 - 76 s t^4.}}
Equation \prolong\ is remarkably simple by comparision with \longlist. That the equations
reduce this way opens up the possibility of answering a host of otherwise intractable 
questions.

\bigbreak\bigskip\bigskip\centerline{{\bf Acknowledgements}}\nobreak
It is our pleasure to thank Robert Bryant for suggesting and explaining deprolongation
which led to the results of section six. 
The work of S.S. is supported  by the William Keck Foundation and by 
NSF grant PHY--9513835; that of M.S. by NSF grant DMS--9870161.

\vfill\eject

\appendix{A}{Quaternions and Symplectic Groups}

We will summarize some useful relations between quaternions and symplectic 
groups.
Let us label a basis for our quaternions by $\{ \1 , I, J, K\}$ where,
$$I^2=J^2=K^2=-\1, \qquad IJK = - \1. $$
A quaternion $q$ can then be expanded in components
$$ q = q^1 + I q^2 + J q^3 + K q^4. $$
The conjugate quaternion $\bar{q}$ has an expansion
$$ q = q^1 - I q^2 - J q^3 - K q^4. $$
The symmetry group $Sp(1)_R \sim SU(2)_R$ is the group of unit quaternions. Let 
$\Lambda $ be a field
transforming in the $\2$ of $Sp(1)_R$. If we view $Sp(1)_R$ acting on $\Lambda$ 
as
right multiplication by a unit quaternion $g$ then,
$$ \Lambda \,\r\, \Lambda g. $$
In this formalism, $\Lambda$ is valued in the quaternions. 
Equivalently, we can expand $\Lambda$ in components and express the action of 
$g$ in the following way,
$$ \Lambda_a \,\r\, g_{ab} \Lambda_b, $$
where $g_{ab}$ implements right multiplication by the unit quaternion $g$.
For example, right multiplication by $I$ on $q$ gives
$$ \eqalign{ q &\,\r\, q I \cr
               & \, \r \,  q^1 I - q^2 - q^3 K + q^4 J,}$$
which can be realized by the matrix
\eqn\defs{  I^R = \pmatrix{0 & -1 & 0 & 0 \cr 
            1 & 0 & 0 & 0 \cr
            0 & 0 & 0 & 1 \cr
            0 & 0 & -1 & 0 }}
acting on $q$ in the usual way $ q_a \, \r\, I^R_{ab} \, q_b$. 
The matrices $J^R$ and $K^R$ realize right multiplication by $J,K$
while ${\1}^R$ is the identity matrix:
\eqn\defstwo{  J^R = \pmatrix{0 & 0 & -1 & 0 \cr 
            0 & 0 & 0 & -1 \cr
            1 & 0 & 0 & 0 \cr
            0 & 1 & 0 & 0 }, \qquad
             K^R = \pmatrix{0 & 0 & 0 & -1 \cr 
            0 & 0 & 1 & 0 \cr
            0 & -1 & 0 & 0 \cr
            1 & 0 & 0 & 0 }. }
We define operators $s^j$ in terms of $\left\{ {\1}^R, I^R, J^R, K^R \right\}$
$$   s^1 = \pmatrix{\1^R & 0 \cr 0 & \1^R}, \quad
                s^2 = \pmatrix{I^R & 0 \cr 0 & I^R}, \quad
                 s^3 = \pmatrix{J^R & 0 \cr 0 & J^R}, \quad
                s^4 = \pmatrix{K^R & 0 \cr 0 & K^R}.$$

In a similar way, the group $Sp(2) \sim Spin(5)$ is the group of 
quaternion-valued
$2\times 2$ matrices with unit determinant. We will
view $Sp(2)$ as acting by left multiplication on a field $\Psi$ in the defining
representation. So an
element $U\in Sp(2)$ acts in the following way:
$$ \Psi \,\r\, U \Psi.$$
Equivalently, in terms of components
$$ \Psi_a \,\r\, U_{ab} \Psi_b. $$   
Lastly, we can give an explicit form for the gamma matrices \gmat\ in terms of 
quaternions:
$$ \g^1 = \pmatrix{1 & 0 \cr 0 & -1 }, \qquad
\g^2 = \pmatrix{0 & 1 \cr 1 & 0 }, \qquad \g^3 = \pmatrix{0 & I \cr -I & 0 } $$
$$ \g^4 = \pmatrix{0 & J \cr -J & 0 }, \qquad \g^5 = \pmatrix{0 & K \cr -K & 0 
}.$$
In turn, $\{ I, J, K \} $ can be expressed in terms of the Pauli matrices $\s^i$
$$ \s^1 = \pmatrix{0 & 1 \cr 1 & 0}, \qquad 
\s^2 = \pmatrix{0 & -i \cr i & 0}, \qquad \s^3 = \pmatrix{1 & 0 \cr 0 & -1} $$ 
as $4\times 4$ real anti-symmetric matrices: 
$$ I = \pmatrix{0 & \s^1 \cr -\s^1 & 0 }, \qquad
 J = \pmatrix{-i\s^2 & 0 \cr 0 & -i \s^2 }, 
\qquad K = \pmatrix{0 & \s^3 \cr -\s^3 & 0 }. $$

\vfill\eject

\appendix{B}{Forms and Representations of $Sp(2)$}

Using the complexification \complexify, we obtain a set of Hermitian $4\times 4$
matrices from those given in Appendix A:
$$ \gg^1 = \pmatrix{1 & 0 \cr 0 & -1 }, \qquad
\gg^2 = \pmatrix{0 & 1 \cr 1 & 0 }, \qquad \gg^3 = \pmatrix{0 & -i \s^1 \cr 
i \s^1 & 0 } $$
$$ \gg^4 = \pmatrix{0 & -i \s^3 \cr i \s^3 & 0 }, 
\qquad \gg^5 = \pmatrix{0 & i \s^2 \cr -i \s^2 & 0 }.$$
We also need the symplectic metric or charge conjugation matrix,
$$ C =  \pmatrix{0 & 1 & 0 & 0 \cr -1 & 0 & 0 & 0 \cr 0 & 0 & 0 & 1 \cr   
0 & 0 & -1 & 0},$$
which implements complex conjugation:
\eqn\conjugate{ C \, \gg \, C = - \gg^*. }

Our forms $du_a$ and $dv_a$ transform in the $\4$ of $Sp(2)$. The representation
$ \wedge^2 \, \4 $ decomposes into $ \5 \oplus \1$. As an example, we can 
explicitly construct the $\1$ from $(1,1)$ forms in the following way, 
\eqn\constructone{ du \, C \, dv, }
while the $\5$ is given by:
\eqn\constructfive{ du \, \gg^\mu \, C \, dv.}  
It is not hard to check that these combinations transform correctly. Lastly,
we need to consider $ \5 \otimes \5  = \1 \oplus {\bf 10} \oplus {\bf 14}$ since
the $\1$ and the ${\bf 14}$ appear in \listiii. The $\1$ is given by the form, 
\eqn\connewone{\sum_\mu du \, \gg^\mu \, C \, du \, dv \, \gg^\mu \, C \, dv,} 
while the ${\bf 14}$ has components:
\eqn\confourteen{ du \, \gg^{(\mu} \, C \, du \, dv \, \gg^{\nu )} \, C \, dv - 
{1\over 5}\, \delta^{\mu\nu} \sum_\rho du \, \gg^\rho \, C \, du \, dv \, 
\gg^\rho \, C \, dv. }

\vfill\eject
\appendix{C}{A Word on Normalizations}

To fix the choice of normalizations, we list explicitly the forms 
appearing in \listi\ through \listxi\ at the special point where  $ x^1 \neq 0$
with $x^\mu =0$ for $\mu>1$ and $ q_1 \neq 0$ with $q_j=0$ for $j>1$. All forms 
act on
the canonical vacuum $|0>$ which is omitted,
\eqn\elisti{
f_1 |0,0> = f_1 (q_1)^2 du_1 du_2 du_3 du_4, }
\eqn\elistii{ \eqalign{
 f_2 |0,4> = & f_2  (q_1)^2  dv_1 dv_2 dv_3 dv_4,  }}
\eqn\elistiii{ \eqalign{
  f_3 |2,2>_{\1} = & f_3 \{ (du_1 du_2-du_3 du_4) (dv_1 dv_2 - dv_3 dv_4)  - \cr 
&
2 ( du_1du_4 dv_2 dv_3 + du_2 du_3 dv_1 dv_4) + \cr & 
2(du_1 du_3 dv_2 dv_4 + du_2 du_4 dv_1 dv_3) \}, \cr   
  f_4 x^1 |2,2>^1_{\5} = & f_4 x^1 (du_1 du_2-du_3 du_4) (dv_1 dv_2 + dv_3 
dv_4), \cr
  f_5 (x^1)^2  |2,2>^{1 1}_{\bf 14}  = & f_5 (x^1)^2 \{ 
{4\over 5} (du_1 du_2-du_3 du_4) (dv_1 dv_2 - dv_3 dv_4) + \cr & {2\over 5} ( 
du_1
du_4 dv_2 dv_3 + du_2 du_3 dv_1 dv_4) - \cr & {2\over 5}
(du_1 du_3 dv_2 dv_4 + du_2 du_4 dv_1 dv_3)\}
,}}
\eqn\elistiv{ \eqalign{
  f_6 |1,3> = & f_6 q_1 (dv_1 dv_2 + dv_3 dv_4)(du_1 dv_2 - du_2 dv_1 + du_3 
dv_4 -
du_4 dv_3), \cr
  f_7 |1,1> = & f_7  q_1 (du_1 du_2 + du_3 du_4) (du_1 dv_2 - du_2 dv_1 + du_3 
dv_4 -
du_4 dv_3), \cr}}
\eqn\elistv{ \eqalign{
  f_8 x^1 |1,3>^1 = & f_8  x^1 q_1 (dv_1 dv_2 + dv_3 dv_4)(du_1 dv_2 - du_2 dv_1 
-
 du_3 dv_4 + du_4 dv_3), \cr
  f_9 x^1 |1,1>^1 = & f_9  x^1 q_1 (du_1 du_2 + du_3 du_4)(du_1 dv_2 - du_2 dv_1 
-
 du_3 dv_4 + du_4 dv_3), \cr}}
\eqn\elistx{\eqalign{ f_{10} |0,2> = & f_{10} \, 
{1\over 2} (q_1)^2 (du_1 du_2 + du_3 du_4)(dv_1 dv_2 + dv_3 dv_4), \cr }}
\eqn\elistxi{\eqalign{ f_{11} x^1 |0,2>^1 = & f_{11} x^1 \, 
{1\over 2} (q_1)^2 (du_1 du_2 + du_3 du_4)(dv_1 dv_2 - dv_3 dv_4). \cr }}

\vfill\eject
\appendix{D}{The Four-Dimensional Radial Harmonic Oscillator}

We want to construct eigenstates for the radial four-dimensional simple harmonic
oscillator which satisfy, 
\eqn\defes{
\left\{ - \partial_y^2 - {3\over y} \partial_y + r^2 y^2 \right\} |n> = E_{n} |n>,}
where $E_n = 4 (n+1) r$. The easiest is the ground state:
$$ |0> = e^{-r y^2/2}. $$
A general eigenstate takes the form,
\eqn\generalform{ |n>  = \left( 1 + a_1^{(n)} y^2 + \ldots + a_n^{(n)} 
y^{2n} \right) e^{-r y^2/2}. }
It is not hard to check using the nice relation, 
\eqn\nicerel{ \left\{ - \partial_y^2 - {3\over y} \partial_y + r^2 y^2 \right\} 
y^{2n} e^{-r y^2/2} =
\left\{ E_n y^{2n} - 4n(n+1) y^{2n-2} \right\} e^{-r y^2/2}, }
that each coefficient $a_m^{(n)}$ is determined by the recursion relation, 
\eqn\recurse{ a_m^{(n)} = {a_{m-1}^{(n)} (m-n-1) r \over m(m+1)},}
where $a_0^{(n)}=1$. Note that these eigenstates are not normalized, but they are orthogonal when
integrated with the measure $ y^3 dy$. The norm of these eigenstates is given by the 
formula,
\eqn\norm{ <n|n> = {1\over r^2} {1\over (2 + 2n)}. }

We will need to evaluate various operators acting on $|n>$. The nicest are the three operators
$y^2, y \partial_y, \partial_r$. These operators raise and lower by at most one unit:
\eqn\relations{ \eqalign{
r y^2 |n> &=  - n |n-1>  + 2(n+1) |n> - (n+2)|n+1>, \cr
y \partial_y |n> &= -n |n-1>  -2 |n> + (n+2) |n+1>, \cr
 2 r \partial_r |n> &= - n |n-1>  -2 |n> + (n+2 )|n+1>. \cr }}
Note that $ y \partial_y$ is equivalent to $ 2 r \partial_r$ when acting on $|n>$. 

\appendix{E}{Equations from a Taylor Expansion}

We list explicitly the equations that follow from a Taylor expansion of the $f_i$ in the 
$y$-direction and which only involve $t_i^0$ and $t_i^2$, 
\eqn\shorttwo{ \eqalign{
(1) & \quad {1\over 2} t_1^0 - {r^2\over 5} t_5^2 + t_{11}^2 + t_3^2 =0, \cr
(2) & \quad r \partial_r t_9^2 -6 t_1^2 - {1\over 2} t_7^0 + 5 t_9^2 + 2 t_{10}^2 =0,\cr
(3) & \quad \partial_r t_7^2 + 2 r t_{11}^2 + r \left\{ t_1^0 - {1\over 2} t_9^0 \right\}
=0, \cr
(4) & \quad 2 r t_9^2 + \partial_r t_1^0 + r t_7^0 =0, \cr 
(5) & \quad 2 t_7^2 + r^2 t_9^0 =0, \cr
(6) & \quad 4 t_{10}^2 - 4 t_1^2 + 5 t_9^2 + r^2 f_{11}^0 + r \partial_r t_9^2 + {1\over 2}
t_7^0 =0, \cr
(7) & \quad 2 r^2 t_4^0 + 3 t_9^0 -r \partial_r t_9^0 + 4 t_3^2 + {16 \over 5} r^2 t_5^2 
=0,\cr
(8) & \quad {4\over 5} r^2 t_5^2 - 4 t_3^2 -r\partial_r t_9^0 -3 t_9^0 =0, \cr
(9) & \quad t_{10}^0 + {1\over 2} t_9^0 + 4 t_{11}^2 + {1\over r} \partial_r t_7^2 =0, \cr
(10) & \quad 2 t_3^0 + {8\over 5} r^2 t_5^0 + 4 t_4^2- {1\over r} \partial_r t_7^0 =0, \cr
(11) & \quad r \left\{ t_4^0 + t_7^0 \right\} - 2 r t_9^2 + \partial_r t_{10}^0 =0, \cr
(12) & \quad {56\over 5} r t_5^2 - r t_7^0 + 6 r t_9^2 + 2 \partial_r t_3^2 + {8\over 5}
r^2 \partial_r t_5^2 =0, \cr
(13) & \quad {2\over r} \partial_r t_3^2 - {2\over 5} r \partial_r t_5^2 - {14\over 5} t_5^2
- t_7^0 - 6 t_9^2 =0, \cr
(14) & \quad 10 t_4^2 + 6 t_7^2 - r^2 t_9^0  + 2 r \partial_r t_4^2 =0, \cr
(15) & \quad t_3^0 + {4\over 5} r^2 t_5^0 + r^2 t_9^0 + t_{11}^0 -2 t_7^2 + r \partial_r 
t_{11}^0 =0.    
}} 
\vfil\eject
\noindent Those equations that only involve $t_i^0, t_i^2$ and $t_i^4$ are given below: 
\eqn\shortfour{\eqalign{
(1) & \quad t_3^4 + {1\over 2} t_1^2 - {r^2\over 5} t_5^4 + t_{11}^4 =0, \cr
(2) & \quad r \partial_r t_9^4 -8 t_1^4 - {1\over 2} t_7^2 + 5 t_9^4 + 2 t_{10}^4 =0,\cr
(3) & \quad \partial_r t_7^4 + 2 r t_{11}^4 + r \left\{ t_1^2 - {1\over 2} t_9^2 \right\}
=0, \cr
(4) & \quad 4 r t_9^4 + \partial_r t_1^2 + r t_7^2 =0, \cr 
(5) & \quad 4 t_7^4 + r^2 t_9^2 + {1 \over 2} t_1^0 =0, \cr
(6) & \quad 6 t_{10}^4 - 4 t_1^4 + 5 t_9^4 + r^2 f_{11}^2 + r \partial_r t_9^4 + {1\over 2}
t_7^2 =0, \cr
(7) & \quad 2 r^2 t_4^2 -{1\over 2} t_7^0 + 3 t_9^2 -r \partial_r t_9^2 + 8 t_3^4 
+ {32 \over 5} r^2 t_5^4 =0,\cr
(8) & \quad {8\over 5} r^2 t_5^4 - 8 t_3^4 -r\partial_r t_9^2 -3 t_9^2 + {1\over 2} t_7^0 =0, \cr
(9) & \quad t_{10}^2 + {1\over 2} t_9^2 + 6 t_{11}^4 + {1\over r} \partial_r t_7^4 =0, \cr
(10) & \quad 2 t_3^2 + {8\over 5} r^2 t_5^2 -{1\over 2} t_9^0 + 8 t_4^4
- {1\over r} \partial_r t_7^2 =0, \cr
(11) & \quad r \left\{ t_4^2 + t_7^2 \right\} - 4 r t_9^4 + \partial_r t_{10}^2 =0, \cr
(12) & \quad {56\over 5} r t_5^4 - r t_7^2 + 8 r t_9^4 + {r\over 2} t_{11}^0 
+ 2 \partial_r t_3^4 + {8\over 5} r^2 \partial_r t_5^4 =0, \cr
(13) & \quad {2\over r} \partial_r t_3^4 - {2\over 5} r \partial_r t_5^4 - {14\over 5} t_5^4
- t_7^2 - 8 t_9^4 =0, \cr
(14) & \quad 10 t_4^4 + 8 t_7^4 - r^2 t_9^2 +{1\over 2} t_{10}^0 + 2 r \partial_r t_4^4 =0, \cr
(15) & \quad t_3^2 + {4\over 5} r^2 t_5^2 + r^2 t_9^2 + t_{11}^2 -4 t_7^4 + r \partial_r 
t_{11}^2 =0.}}

\listrefs
\bye